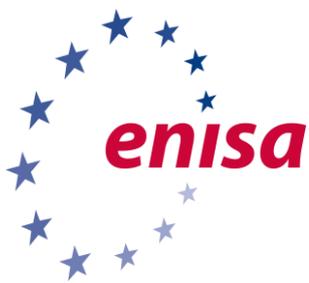

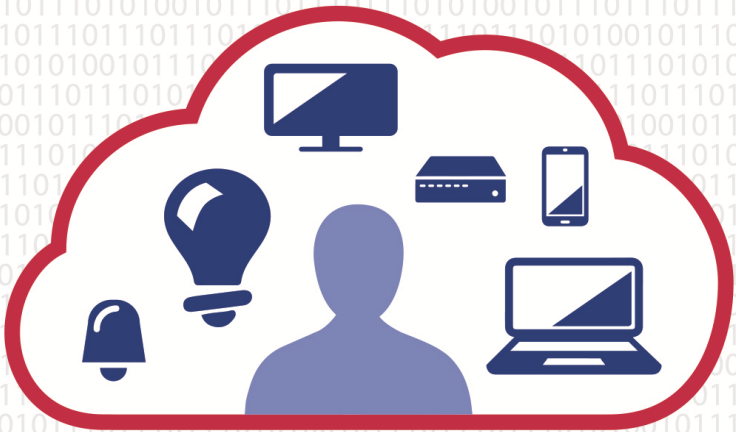

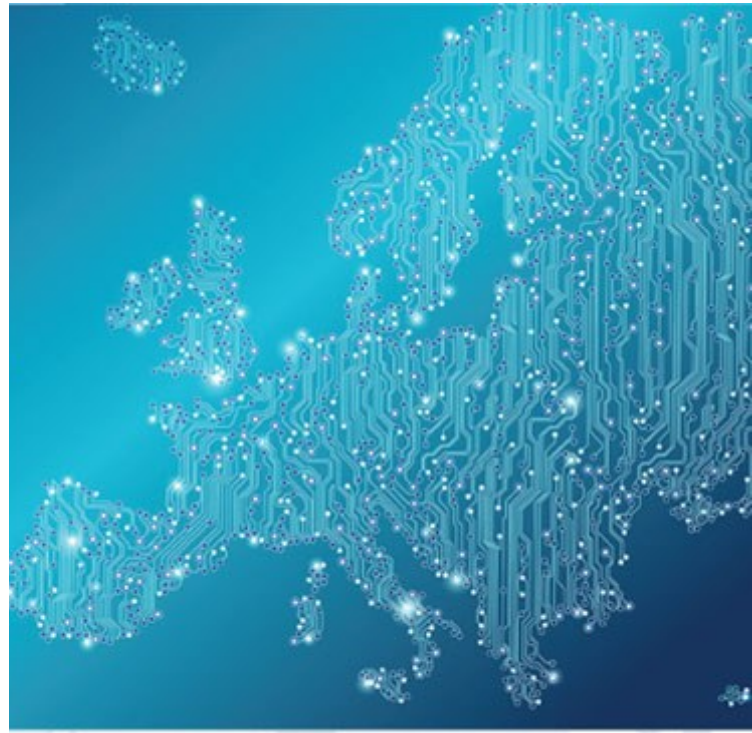

# Privacy by design in big data

An overview of privacy enhancing technologies in the era of big data analytics

FINAL
1.0
PUBLIC
DECEMBER 2015

www.enisa.europa.eu          European Union Agency For Network And Information Security

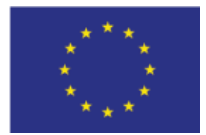

# About ENISA

The European Union Agency for Network and Information Security (ENISA) is a centre of network and information security expertise for the EU, its member states, the private sector and Europe's citizens. ENISA works with these groups to develop advice and recommendations on good practice in information security. It assists EU member states in implementing relevant EU legislation and works to improve the resilience of Europe's critical information infrastructure and networks. ENISA seeks to enhance existing expertise in EU member states by supporting the development of cross-border communities committed to improving network and information security throughout the EU. More information about ENISA and its work can be found at www.enisa.europa.eu.

## Authors
**Giuseppe D' Acquisto** (Garante per la protezione dei dati personali), **Josep Domingo-Ferrer** (Universitat Rovira i Virgili), **Panayiotis Kikiras** (AGT), **Vicenç Torra** (University of Skövde), **Yves-Alexandre de Montjoye** (MIT), **Athena Bourka** (ENISA)

## Editors
European Union Agency for Network and Information Security
ENISA responsible officers: **Athena Bourka, Prokopios Drogkaris**
For contacting the authors please use isdp@enisa.europa.eu.
For media enquiries about this paper, please use press@enisa.europa.eu.

## Acknowledgements
We would like to thank **Gwendal Le Grand** (CNIL) for his support and advice during the project. Acknowledgements should also be given to **Stefan Schiffner** (ENISA) for his help and support in producing this document.

# Table of Contents

# Executive Summary

The extensive collection and further processing of personal information in the context of big data analytics has given rise to serious privacy concerns, especially relating to wide scale electronic surveillance, profiling, and disclosure of private data. In order to allow for all the benefits of analytics without invading individuals' private sphere, it is of utmost importance to draw the limits of big data processing and integrate the appropriate data protection safeguards in the core of the analytics value chain.  ENISA, with the current report, aims at supporting this approach, taking the position that, with respect to the underlying legal obligations, the challenges of technology (for big data) should be addressed by the opportunities of technology (for privacy).

To this end, in the present study we first explain the need to shift the discussion from "big data versus privacy" to "big data with privacy", adopting the privacy and data protection principles as an essential value of big data, not only for the benefit of the individuals, but also for the very prosperity of big data analytics. In this respect, the concept of privacy by design is key in identifying the privacy requirements early at the big data analytics value chain and in subsequently implementing the necessary technical and organizational measures.

Therefore, after an analysis of the proposed privacy by design strategies in the different phases of the big data value chain, we provide an overview of specific identified privacy enhancing technologies that we find of special interest for the current and future big data landscape. In particular, we discuss anonymization, the "traditional" analytics technique, the emerging area of encrypted search and privacy preserving computations, granular access control mechanisms, policy enforcement and accountability, as well as data provenance issues. Moreover, new transparency and access tools in big data are explored, together with techniques for user empowerment and control.

Following the aforementioned work, one immediate conclusion that can be derived is that achieving "big data with privacy" is not an easy task and a lot of research and implementation is still needed. Yet, we find that this task can be possible, as long as all the involved stakeholders take the necessary steps to integrate privacy and data protection safeguards in the heart of big data, by design and by default. To this end, ENISA makes the following recommendations:

- **Privacy by design applied**

In this report we tried to explore the privacy by design strategies in the different phases of the big data value chain. However, every case is different and further guidance is required especially when many players are involved and privacy can be compromised in various points of the chain.

*Data Protection Authorities, data controllers and the big data analytics industry need to actively interact in order to define how privacy by design can be practically implemented (and demonstrated) in the area of big data analytics, including relevant support processes and tools.*

- **Decentralised versus centralised data analytics**

Big data today is about obtaining and maximizing information. This is its power and at the same time its problem. Following the current technological developments, we take the view that selectiveness (for effectiveness) should be the new era of analytics. This translates to securely accessing only the information that is actually needed for a particular analysis (instead of collecting all possible data to feed the analysis).

*The research community and the big data analytics industry need to continue and combine their efforts towards decentralised privacy preserving analytics models. The policy makers need to encourage and promote such efforts, both at research and at implementation levels.*

- **Support and automation of policy enforcement**

In the chain of big data co-controllership and information sharing, certain privacy requirements of one controller might not be respected by another. In the same way, privacy preferences of the data subjects may also be neglected or not adequately considered. Therefore, there is need for automated policy definition and enforcement, in a way that one party cannot refuse to honour the policy of another party in the chain of big data analytics. Semantics and relevant standards, as well as cryptographically enforced rules, are fields that require careful study in this respect.

*The research community and the big data analytics industry need to explore the area of policy definition and to embody relevant mechanisms for automated enforcement of privacy requirements and preferences. The policy makers need to facilitate the dialogue between research, industry and Data Protection Authorities for effective policy development and automation models.*

- **Transparency and control**

In the age of big data, "traditional" notice and consent mechanisms fail to provide proper transparency and control. In the context of this report, we discussed some interesting concepts and ideas, such as the privacy icons, sticky policies and personal data stores. Still, the very idea of consent needs to be reinforced with new models and automated enforcement mechanisms.

*The big data analytics industry and the data controllers need to work on new transparency and control measures, putting the individuals in charge of the processing of their data. Data Protection Authorities need to support these efforts, encouraging the implementation of practical use cases and effective examples of transparency and control mechanisms that are compatible with legal obligations.*

- **User awareness and promotion of PETs**

There are already numerous privacy enhancing tools for online and mobile protection, such as anti-tracking, encryption, secure file sharing and secure communication tools, which could offer valuable support in avoiding unwanted processing of personal data. Still, more work is needed in evaluating the reliability of these tools and their further applicability for the general public.

*The research community needs to adequately address aspects related to the reliability and usability of online PETs. The role of the Data Protection Authorities is central in user awareness and promotion of privacy preserving processes and tools in online and mobile applications.*

- **A coherent approach towards privacy and big data**

This report focused mainly on technology for big data privacy. Still technology alone is not enough. The legal obligations for data protection in EU need to be practically implemented, also in relation to the overall European Commission's vision and research programme for big data. To this end, big data research should include as a core element the integration of privacy enhancing technologies. And research on privacy enhancing technologies should take into account the dimension and emerging big data landscape.

*Policy makers need to approach privacy and data protection principles (and technologies) as a core aspect of big data projects and relevant decision-making processes.*

Finally, taking into account the continuously growing technological environment, this report should not be considered as an exhaustive presentation of all available privacy enhancing techniques, but rather as an attempt to take the discussion of "big data with privacy" a step forward, following a technological perspective. ENISA aims at continuing its work in this field, based on a multi-disciplinary approach and facilitating the dialogue between all involved parties.

# 1. Introduction

During the last few years, big data has evolved to an emerging field where innovation on technology allows for new ways to deal with huge amounts of data created in near real time by a vast variety of sources (IoT sensing devices, M2M communication, social applications, mobile video, etc.). Big data can provide for "big" analytics that offer novel opportunities to reuse and extract value from the "information chaos", escaping the confines of structured databases, identifying correlations, conceiving new, unanticipated uses of data. Big analytics can offer a whole new area of opportunities from research to online transactions and service provision in several sectors of everyday life. This has been recognised by the European Commission, which in its latest Communication on big data stresses the need for a data-driven economy, contributing to citizens' welfare and socio-economic growth [1].

Nevertheless, while these efforts are in many cases welcomed, for example when they support prediction of climate change, epidemics spread or side effects of medicines, they can often pose serious challenges to privacy and the protection of personal data. This danger has been widely argued and outlined by the EU privacy community [2][3][4] and has even led the European Parliament, in its latest study on big data, to contest the very idea of a data-driven economy and "datafication" of society [5].

Despite the benefits of analytics, it cannot be accepted that big data comes at a cost for privacy. At the same time technology and innovation cannot be stopped. It is, thus, of utmost importance to craft the right balance between making use of big data technologies and protecting individuals' privacy and personal data. This report focuses exactly on striving this balance by highlighting privacy as a core value of big data and examining how technology can be on its side. In particular, following ENISA's former work on privacy and data protection by design [6], we aim at contributing to the big data discussions by defining privacy by design strategies and relevant privacy enhancing technologies, which can allow for all the benefits of analytics without compromising the protection of personal data.

To this end, we first present in Chapter 2 the need to shift the discussion from "big data versus privacy" to "big data with privacy", explaining why the new privacy risks in the big data era should be managed for the prosperity of both the individuals and the big data analytics industry, thus for the society as a whole. In this context we also show how the privacy principles enshrined in the EU legal framework for the protection of personal data can contribute to building trust in big data, thereby providing for a more humanistic technological development.

Chapter 3 then proposes privacy by design as a critical tool for addressing big data risks. Starting from the basic concept and definition, we analyse the particular privacy by design strategies in each phase of big data analytics, together with the subsequent technical implementation measures.

Based on the above-mentioned description, Chapter 4 provides an overview of the main privacy enhancing technologies that are promising for big data. Having said that, it is important to note that it is not really about new technologies but rather about the specificities of existing technologies in the big data environment. Special focus is put on big data anonymization, encryption, privacy by security, transparency, access and control mechanisms. The overall objective is to show the state-of-the-art regarding the use of these technologies in the context of big data implementations, as well as examine current research and trends in this field.

Chapter 5, drawing on the previous chapters, includes conclusions and recommendations on the technical integration of a privacy by design approach in big data architectures, as well as on the policy requirements for supporting such an approach.

Annex 1 shows an example of big data analytics in practice in the area of smart cities. In particular, we present three specific big data use cases in the context of a smart city (smart parking, smart grids, citizen platform) and we further discuss the privacy risks and corresponding mitigation measures.

Annex 2 provides some background information on anonymization (supporting the topics addressed in the main text).

This document provides information on privacy and big data for all interested stakeholders, such as developers and industry of big data analytics, data controllers making use of big data technologies, as well as data protection regulators and supervisors at EU and Member States levels.

ENISA in 2015 published two more documents in the area of big data[1]. The first one is the "Big data threat landscape and good practice guide", which provides an overview of current and emerging threats applicable to big data technologies, together with their associated trends. The second is on "Big data security" which provides good practices and recommendations on the security of big data systems. These documents and the present study aim at complementing each other in providing a comprehensive and multi-dimensional overview of the current big data security and privacy landscape.

---

[1] See latest publications at ENISA's web site, https://www.enisa.europa.eu/

# 2. From "big data versus privacy" to "big data with privacy"

When dealing with privacy issues in the current debate surrounding big data analytics, the impression that one may often gather is that we are in the presence of a conflict between positions which cannot easily be reconciled. It is as if privacy were an obstacle to innovation represented by the opportunities offered by big data, a burden coming from the past. Or as if big data will bring the end of privacy, an inevitable violation of the private sphere for the sake of technological innovation. We tend to be skeptical on this idea of a conflict and, instead, we think that this is just history repeating itself, like every time a technology shift happens, especially at its early stage. At the end, it is all a naïve confrontation between those who only see the dangers and those who only perceive the benefits[2]. The story however is much more complex and, over the time, these requirements cannot necessarily fit in stereotypical schemes. To state it differently: big data analytics are here to stay, as well as privacy. The goal of this chapter is to outline how the technological capability to extract value from data for modern society and the control over it, which is embodied by privacy principles, can (and will) prosper together.

## 2.1 Understanding big data

Business consultants Gartner Inc. define big data as "high-volume, high-velocity and high-variety information assets that demand cost-effective, innovative forms of information processing for enhanced insight and decision making" [7]. This definition points out the three most outlined dimensions of big data (also known as the 3Vs[3] that define big data):

- **Volume**: Huge amounts of data in the scale of zettabytes [4] and more.

*Today, Facebook alone ingests 500 terabytes of new data per day[5]. According to IBM every day 2.5 quintillions of bytes of data are generated: this sums up to 4 zettabytes of data worldwide in 2013 [8] and further enlargement is expected with the 50 billion of sensors going online by 2020.*

- **Velocity**: Real time streams of data flowing from diverse resources (e.g. physical sensors or "virtual sensors" from social media, such as Twitter streams).

*For example, online data can be harvested and captured at millions of events per second. Analytics algorithms can provide market trends and predict user behaviour in microseconds. Sensors continuously generate massive production of logs. These are only a few examples of analytics occurring almost in real time today.*

- **Variety**: Data from a vast range of systems and sensors, in different formats and datatypes.

*Numerical data, categorical data, geospatial data, 3D data, audio and video, unstructured text, including log files and social media, all form part of the big data ecosystem.*

---

[2] Umberto Eco, *Apocalittici e integrati: comunicazioni di massa e teorie della cultura di massa*, Milano, Bompiani, 1964 (in english, *Apocalypse Postponed*, John Wiley & Sons 1994).
[3] The first one that talked about the V's was Doug Laney in 2001 [227] who described the "3Vs"—volume, variety, and velocity—as the key "data management challenges" for enterprises. The same "3Vs", that have been used and extended to 4Vs, 5Vs or even 6Vs, are used by almost anyone attempting to define or describe big data.
[4] 1 ZB = $1000^7$ bytes = $10^{21}$ bytes.
[5] http://www.cnet.com/news/facebook-processes-more-than-500-tb-of-data-daily/

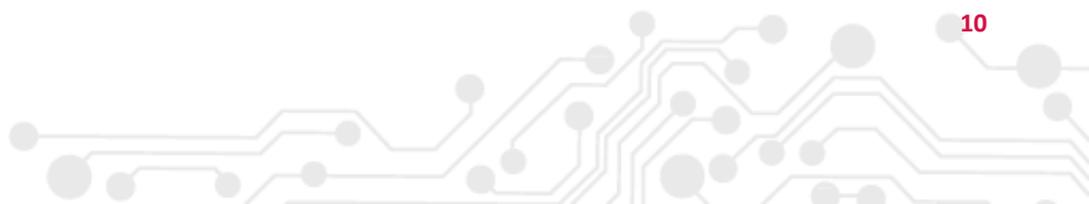

For the purpose of this report one more interesting dimension of big data is also veracity, which describes the incompleteness (inconsistency, inaccuracy) of data.

Due to the above-mentioned characteristics, big data is seen today as the new opportunity for analytics that can offer significant advancements to several aspects of our everyday life, including health, leisure, environment, employment, etc. To this end, data has been characterized by many as the fuel of the 21$^{st}$ century economy or the new oil[6]. Still, data differ from oil in one critical element: they are not just an available (and rather difficult to find) resource but, on the contrary, they are constantly generated by people's activities. *This is why big data may (and in many cases do) also involve personal data, for example a name, a picture, contact details, posts on social networking websites, healthcare data, location data or a computer IP address.*

### 2.1.1 Big data analytics

The term "big data analytics" refers to the whole data management lifecycle of collecting, organizing and analysing data to discover patterns, to infer situations or states, to predict and to understand behaviours. Its value chain includes a number of phases that can be summarised as follows[7]:

- **Data Acquisition/Collection:** the process of gathering, filtering and cleaning data before it is put in a data repository or any other storage solution on which data analysis can be carried out. Examples of potential sources are social networks, mobile apps, wearable devices, smart grids, online retail services, public registers, etc. As the main purpose is to maximise the amount of available data (so as to appropriately feed the analysis), the process is usually based on fast and massive data collection, thus, assuming high-volume, high-velocity, high-variety, and high-veracity but low-value data.

- **Data Analysis:** the process concerned with making the "raw" collected data amenable for decision-making as well as domain specific usage. The key challenge of data analysis is to find what is really useful. A critical element in that respect is to combine data from different sources in order to derive information that cannot be found otherwise. Data analysis covers structured or unstructured data, with/without semantic information and can have multiple levels of processing and different techniques (e.g. diagnostic, descriptive, predictive and prescriptive analysis).

- **Data Curation:** the active management of data over its lifecycle to ensure it meets the necessary quality requirements for effective usage. It includes functions like content creation, selection, classification, transformation, validation and preservation. A main aspect in that respect is the need to assure the reusability of the data, not only within their original context but in many different contexts.

- **Data Storage:** storing and managing data in a scalable way satisfying the needs of applications/analytics that require access to the data. Cloud storage is the trend but in many cases distributed storage solutions would be best options (e.g. for streaming data).

- **Data Usage:** covers the use of the data by interested parties and is very much dependent on the data processing scenario. For example the results from an analysis on trends in mobile apps usage could be available for the general public or restricted to a mobile service provider who commissioned the study. Therefore, users of big data may vary from a single organisation to a wide range of parties, such as banks, retailers, advertising networks, public authorities, etc.

---

[6] See for example, http://www.wired.com/insights/2014/07/data-new-oil-digital-economy/
[7] See also the EU project "Big Data Public Private Forum" for a thorough analysis of the big data value chain, http://www.big-project.eu/

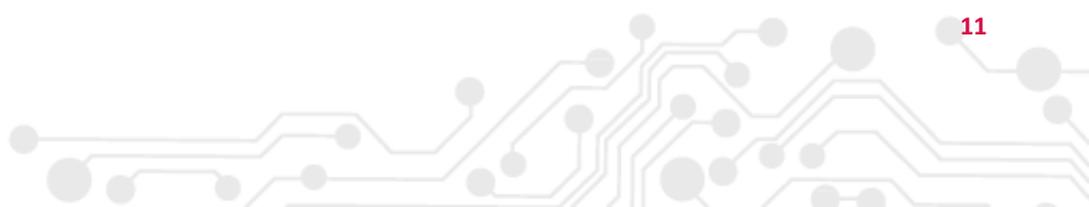

Big data analytics is already happening today. Many activities of our everyday life pertain analytics, which we fuel in return during some of our everyday transactions. Some usual examples include:

- In hospitals, patients' vital signs can be compared against historical data to identify patterns and provide valuable information for early detection and treatment of diseases.
- Mobile fitness apps collect data on the way we walk, sleeping patterns or other streams of data, which if combined with health data, can help healthcare providers offer better wellness programs.
- Companies offer smart home functionality that enables motion detection to monitor what is happening in one's home and remotely control devices via a smartphone app.
- During our everyday travelling through the city, location information is sent by our smartphone regarding where we are and how fast we move, in order to get information about nearby points of interest.
- Google's self-driving car is analysing huge amounts of data from sensors and cameras in real time to stay safely on the road[8].
- Smart TVs can track what we watch and provide accordingly suggestions or even ads based on our preferences[9].

As it is obvious from the examples, a variety of stakeholders are involved in the different phases of the big data value chain, including hardware, software and operating system vendors, different types of service providers (telecom operators, social networks, cloud providers, etc.), analytics providers, data brokers, public authorities, etc. These stakeholders may take different roles during a big data analytics scenario and interact with each other in variant ways.

## 2.2 Privacy and big data: what is different today?

Analysing data in order to support decision making, discover trends or open new business opportunities is not new. Neither is the collision of certain types of such processing with privacy and data protection principles. The EU legal framework on data protection is aimed exactly at dealing with this collision by regulating the boundaries within which personal data can be processed with respect to the individuals' private life and information. So, what is new about big data?

Simply put: it is the scale of big data processing which brings existing privacy risks into a whole new (and unpredictable) level. The scale is in terms of volume, variety, velocity and veracity, all the Vs of the big data definition, and their combination in analytics technologies. To this end, the main privacy challenges associated with big data, overrunning the privacy concerns of "traditional" data processing, are as follows:

---

[8] See: https://www.google.com/selfdrivingcar/
[9] See for example, http://www.theverge.com/2015/2/11/8017771/samsung-smart-tvs-inserting-unwanted-ads

- **Lack of control and transparency**

The collection of big data is based on so many different and unexpected sources that control by the individual can easily be lost, as in many cases he/she is not even aware of the processing or cannot trace how data are flowing from one system to another. This poses significant engineering challenges on how to effectively and timely inform users and on who is in charge of this task, especially when the processing requires the interaction of many players.

*Examples of such unknown and/or uncontrolled processing include data captured from sensors and cameras, from screening posts in social networks or from analysis of web searches.*

- **Data reusability**

One of the main targets of big data analytics is to use data, alone or in combination with other data sets, beyond their original point and scope of collection. The scalability of storage allows for potential infinite space, which means that data can be collected continuously until a new value can be created from insights derived out of them.

*For example, mobile apps providers collect personal data in order to provide users with information about their fitness or health status (e.g. heart condition, dietary habits, etc.). These data can be valuable to insurance companies and/or other providers (e.g. sports centres, dietary consultants, etc.) who may target specific users.*

- **Data inference and re-identification**

Another important element for big data is the possibility to combine data sets from many different sources, so as to derive more (and new) information. This also triggers privacy risks, especially if linking different sources may allow the emergence of patterns related to single individuals. For example, it is feasible by combining various allegedly non-personal data to infer information related to a person or a group [9]. If the likelihood of such inference is high, it may be the case that an unintended processing of personal data occurs. In the same way, the combination of anonymised data sets and advanced analytics can lead to re-identification of a person by extracting and combining different pieces of information [10] [11]. Data inference and re-identification may in certain cases seriously affect an individual's private life, for example leading to humiliation or even threat to life due to the disclosure of confidential information.

*The AOL data breach already in 2006 was one of the first widely known cases of re-identification: the company's release of users' search logs allowed singling out certain users based on their searches* [12] [13]. *The Netflix case followed in 2007 when researchers identified individual users by matching data sets with film ratings on the Internet Movie Database* [14]. *In 2012 the retail company Target, using behavioural advertising techniques, managed to identify a pregnant teen girl from her web searches and sent her relevant vouchers at home[10].*

---

[10] See: http://www.forbes.com/sites/kashmirhill/2012/02/16/how-target-figured-out-a-teen-girl-was-pregnant-before-her-father-did/

- **Profiling and automated decision making**

The analytics applied to combined data sets aim at building specific profiles for individuals that can be used in the context of automated decision making systems, e.g. for offering or excluding from specific services and products. Such a profiling can in certain cases lead to isolation and/or discrimination, including price differentiation, without providing the individuals the possibility to contest these decisions. Moreover, profiling, when based on incomplete data, can lead to false negatives, unfairly depriving individuals from rights and benefits that they are entitled to.

*One evident example of profiling today is the whole area of online behavioural advertising, which is also closely related to price differentiation (perceived by the willingness of people to pay)* [15]. *Another example is that of profiling in predictive police algorithms, which can lead to discrimination of people, even based on the area where they live[11].*

On top of the risks mentioned above, one additional challenge in big data is the difficulty in enforcing, and/or monitoring the data protection controls. These difficulties are similar to the ones of cloud computing, for example in relation to the location of the data and the possibility to conduct audits. Still, in big data there is one additional and critical element: the involvement and interaction of many diverse stakeholders, which makes it even more difficult (for regulators, data controllers and users) to identify privacy flaws and impose relevant measures.

Therefore, the new thing in big data is not the analytics itself or the processing of personal data. It is rather the new, overwhelming and increasing possibilities of the technology in applying advanced types of analyses to huge amounts of continuously produced data of diverse nature and from diverse sources. The data protection principles are the same. But the privacy challenges follow the scale of big data and grow together with the technological capabilities of the analytics.

## 2.3 The EU legal framework for the protection of personal data

The current relevant legal framework to assess privacy issues raised by big data analytics in the EU is composed of Directive 95/46/EC and Directive 2002/58/EC (as amended by Directive 2009/136/EC). The conditions of its applicability are set forth in article 4 of Directive 95/46/EC, and in particular this framework applies to all processing of personal data carried out "in the context of an establishment" of the controller on the territory of a Member State or, in case the controller is not established on Community territory, if it makes use of "equipment" situated on the territory of that Member State[12].

EU law applies to the processing of personal data as defined in article 2 of Directive 95/46/EC, namely to any information relating to an identified or identifiable natural person [16]. In the context of big data analytics, the focus is more on *indirect identification*, and it has to be stressed that processing of personal data may take place whenever it is possible to isolate some or all records which identify an individual in the data set (singling out of a record), or to link, at least, two records concerning the same individual in the same database or in two different databases (linkability of two records), or it is possible to deduce, with significant probability, the value of an attribute from the values of a set of other attributes (inference of attributes)

---

[11] See http://theconversation.com/big-data-algorithms-can-discriminate-and-its-not-clear-what-to-do-about-it-45849
[12] The European Court of Justice, in its judgement of 13 May 2014, Case C-131/12, ruled that the EU legal framework is applicable also for cases where, even if the physical server of a company processing data is located outside Europe, the controller has a branch or a subsidiary in a Member State which promotes the revenues coming from the processing.

[17]. Having said that, it is important to note that individuals can be singled out in many circumstances, ranging from reverse engineering of bad pseudonyms, to the identification of unique mobility patterns, or combinations of attributes, tastes, habits, and even their writing style.

Directive 95/46/EC also refers to cases where the processing of data may fall outside the EU legal framework in Recital 26, which signifies that this is possible whenever the data are stripped of sufficient elements such that the data subject can no longer be identified, a technique known as *anonymization*. More precisely, data must be processed in such a way that it can no longer be used to identify a natural person by using "all the means likely reasonably to be used". An important point to make is that the Directive does not mandate how such a de-identification process should or could be performed. The focus is on the outcome: data should be such as not to allow the data subject to be directly or indirectly identified via "all" "likely" and "reasonable" means.

The concept of data controller is crucial in the application of the EU legal framework on data protection to the big data context. A controller, according to article 2(d) of the Directive 95/46/EC, plays a very central role since this is the entity which "determines the purposes and the means of the processing of personal data". Due to these responsibilities, data controllers have specific obligations.

The principles relating to data quality, contained in article 6 of Directive 95/46/EC, constitute a very important cornerstone of EU data protection law. First of all, personal data should be collected and processed fairly and lawfully. The *fairness principle* requires that personal data should never be processed without the individual being actually aware of it. Moreover, the *purpose limitation principle* implies that data can only be collected for specified, explicit and legitimate purposes. These purposes must be defined before the data processing takes place, and any further purpose incompatible with these original purposes would be illicit under EU law. Then, the data processed should be strictly the ones which are necessary for the specific purpose previously determined by the data controller. This is the *data minimization principle*. Also, article 6 requires that processed personal data are kept for a period which should be no longer than necessary for the purpose for which the data were collected, after which the data should be deleted. Anonymization of data is another option available today to big data stakeholders.

Also, in order for the processing of personal data to be legitimate, data controllers need to fulfil one of the requirements listed in article 7 of Directive 95/46/EC. In practice, the primary legal bases which may be relevant in the context of big data analytics are *consent*, the accomplishment of *contractual obligations* and *legitimate interest*. Consent is the first referred legal basis (article 7(a)) that should be relied on, and in order to be valid it has to be freely given (the data subject must have the choice to accept or refuse the processing of his/her personal data), informed (the data subject must have the necessary information at his/her disposal in order to form an accurate judgement), specific (the expression of will must relate to the purpose for which data are processed) and unambiguous (a positive action signalling the data subject's wish is expected before the processing takes place). Article 7(b) also provides that the processing is legitimate for the performance of a contract, insofar there is a direct and objective link between the processing and the purposes of the contractual performance expected from the data subject. Thirdly, article 7(f) permits the processing of personal data where it is necessary for the purposes of the legitimate interests pursued by the controller or by the third party or parties to whom the data are disclosed, except where such interests are overridden by the interests or fundamental rights and freedoms of the data subject.

Another relevant piece of legislation for big data analytics is art. 5(3) of Directive 2002/58/EC, which is applicable to situations when a big data stakeholder stores or gains access to information already stored on a user's device. This provision demands that the data subject concerned consents to such storage or access for these actions to be legitimate, unless they are "strictly necessary in order to provide a service explicitly requested by the subscriber or user". In addition, this requirement is particularly important in the big data context, since a dispersed number of stakeholders may take part in the provision of a service. The consent requirement, in fact, equally applies to any stakeholder that wants to have access to data stored in the user's terminal.

Specific *transparency requirements* weigh on data controllers, in application of article 10 of Directive 95/46/EC. Namely, on the identity of the controller, the purposes of the processing, the recipients of the data, the existence of their rights of access and right to object. Availability and clarity of this information is a prerequisite for the validity of data subjects' consent, and it is a very important tool to re-balance any asymmetry which may occur in the relationship between users and service providers. These aspects are particularly relevant for big data analytics, when the reuse of data is at stake.

Article 17 of the Data Protection Directive provides that any stakeholder engaged in big data analytics processing that qualifies as a data controller is fully responsible for the *security of the data processing*, with the obligation of implementing appropriate technical and organizational measures to protect personal data. Also, where processing is carried out by a third party, of choosing a processor providing sufficient guarantees in respect of the technical and organizational security measures governing the processing to be carried out. In the context of big data analytics this mainly means implementing appropriate controls to limit access to data, especially taking stock of the effects of correlation, which is the distinctive feature of big data analytics.

*Data subjects have rights* also in a big data environment, as laid down in article 12 (right of access) and article 14 (right to object) of Directive 95/46/EC. In particular, any data subject is entitled to obtain from the data controllers communication in an intelligible form of the data that is subjected to processing and to know the logic involved in any automatic processing of data concerning him/her. These two provisions are particularly relevant in the context of big data analytics, also as a means to limit lock-ins and other competition impediments and to enhance transparency and trust between users and service providers. In addition, data subjects have the right to revoke any prior consent and to object to the processing of data relating to them. The exercise of such rights should not be burdensome for the data subject and withdrawal schemes should be granular, meaning that they should apply potentially to any selection of processed data or purposes.

### 2.3.1 The proposed General Data Protection Regulation

It is important to note that Directive 95/46/EC is at the moment of this report undergoing a significant reform, in particular the Commission proposal for a General Data Protection Regulation[13]. The discussions on the Regulation are actually in their final stage (trilogue discussions between EU Council, Parliament and EC) and the adoption of the Regulation is expected very soon. Generally speaking, the Regulation does not change the core data protection principles of Directive 95/46/EC, but refines them in order to cope with the new technological context, introducing also more safeguards for data subjects and new users' rights. In this respect, there are some interesting new elements to point out in relation to big data.

As a first point, the Regulation places a remarkable emphasis on reinforced and actual transparency (article 14). New elements are in fact included in the list of items on which the data subject must be informed regarding the processing, namely the contact details of a responsible person, the legal basis of the processing (including the specific legitimate interests pursued by the controller or by a third party), as well as the existence of the right to *data portability*. The latter is a prominent new right (article 18), introduced with the scope of reinforcing users' access right. A data subject can in fact receive the personal data concerning him or her, which he or she has provided to a controller, in a structured and commonly used and machine-readable format and have the right to transmit those data to another controller. Also, the right to delete will be reinforced with the introduction of the so called *right-to-be-forgotten* (article 17), aiming at fostering a more responsible and cautious attitude towards public disclosure of data[14]. In particular, this new right envisages obligation for a controller who has made personal data public to take reasonable steps to inform other controllers processing the data, that the data subject has requested the erasure of any links to, or copy or replication of the personal data.

---

[13] See http://ec.europa.eu/justice/data-protection/index_en.htm for more information on the EC proposal and the current status of the data protection reform.

The Regulation introduces also a new awareness tool for data controllers (article 33) which should facilitate compliance with data protection obligations, and which is relevant in particular when using new technologies: the *Data Protection Impact Assessment* (DPIA). This assessment shall contain a description of the envisaged processing operations, an evaluation of the privacy risks and the measures envisaged to address those risks. Moreover, the Regulation introduces for the first time an obligation to the data controller for *data protection by design and by default* (article 23), as a tool to integrate data protection safeguards early at the development stage of new products and services.

Interestingly, new mechanisms for co-controllership are introduced (article 24), with the scope of coping with the complexities of modern data value chain. Again, where two or more controllers jointly determine the purposes and means of the processing of personal data, it is envisaged that they shall transparently determine their respective responsibilities for compliance, designating which of the joint controllers shall act as single point of contact for data subjects to exercise their rights.

It is just worth mentioning that the same approach to foster the implementations of these new or reinforced safeguards, is evident in the definition of economic incentive-deterrence mechanisms, such as the one envisaged for compensation sharing of the damages suffered by data subjects (article 77). Translating the complexity of the processing into the simplicity of the economic transactions between the stakeholders is (this seems to be the stand of the Regulation) a powerful complementary tool to enhance data subjects' privacy, and the issue of joint liability will deserve an important focus in the years to come[15].

## 2.4 Privacy as a core value of big data

Comparing the challenges that privacy faces in the era of big data against the data protection principles enshrined in the EU legal framework, one could at first sight conclude that big data and privacy are in many cases contradictory. The analytics' need for data reuse goes against the principle of purpose limitation, the requirement for massive data collection opposes data minimization, the involvement of many controllers and the complicated interaction between them makes transparency and control very difficult, and so on. Still, as already mentioned, it is not our purpose in this chapter to talk about a clash between big data and privacy. On the contrary what we want to point out is that there is no big data without privacy. It is not really a matter of clash, but a matter of prosperity: if privacy principles are not respected, big data will fail to meet individuals' needs; if privacy enforcement ignores the potential of big data, individuals will not be adequately protected. Therefore, all involved stakeholders should work together in addressing the new challenges and highlighting privacy as a core value of big data. Technology, instead of being a rival in this attempt, should be the main weapon and support tool.

---

[14] For further reading on the topic, International Working Group on Data Protection in Telecommunications (IWGDPT), Working Paper and Recommendations on the Publication of Personal Data on the Web, Website Contents Indexing and the Protection of Privacy (15./16. April 2013, Prague) and ENISA, The right to be forgotten - between expectations and practice (2012), https://www.enisa.europa.eu/activities/identity-and-trust/library/deliverables/the-right-to-be-forgotten

[15] See Maurizio Naldi, Marta Flamini, Giuseppe D'Acquisto, Information Security Investments: When Being Idle Equals Negligence, in Economics of Grids, Clouds, Systems, and Services, Lecture Notes in Computer Science Volume 8193, 2013, pp 268-279 and Solon Barocas and Andrew Selbst, Big Data's Disparate Impact, California Law Review, Vol. 104: 2016 available at http://papers.ssrn.com/sol3/papers.cfm?abstract_id=2477899

### 2.4.1 The oxymoron of big data and privacy

Before analyzing in more detail the (tight) interrelationship between big data and privacy, we would like to take a step back and image the world of "big data without privacy", i.e. massive spread of analytics without particular controls for the protection of personal data.

Aside from privacy considerations, it is easy to conclude that such a world would actually lead to what we might define commoditization [18] of personal data. Namely, a wide availability of data characterized by an increasing level of homogeneity and interoperability, easily machine readable, plug-&-play for any sort of algorithmic forecast. In such a scenario, data referring to identified or identifiable persons, insofar they are distinguishable one from the other in terms of attributes, would not be any longer a scarce resource.

But if personal data were so widely available, and their fully fledged variety constrained in a, even if apparently wide, nevertheless limited number of homogeneous classes, without any form of regulatory protection of their scarcity, their informational value would be lower. For example, people would gradually start being more reluctant in providing their data or they would provide false data[16], so as to get the services they want without discovering their identity. Therefore, over the time the data would become of decreasing added value to a decision process, triggering a rush to the bottom. In commoditized environments, there is little incentive to invest in innovation and the cheapest item is the most attractive one. This would pave the way to a severe reduction of data quality.

This seems to be an oxymoron when applied to big data and privacy: personal data retain their value as long as they are perceived (and they are) a scarce and difficult to obtain resource, which does effectively and differently impact each individual's life. It is not necessarily in the interest of the big data industry and of the individuals to dilute this value. On the contrary, the push in the opposite direction of de-commoditization of personal data (that is, higher value and lower availability), even if not out rightly evident yet, is a major force coming from the individuals who generate the processed data[17]. The big data analytics industry will ultimately have to cope with this challenge, as the scarcity of personal data should be seen as a value to protect, which is in the interest of the big data analytics service providers.

### 2.4.2 Privacy building trust in big data

Following the above-mentioned oxymoron, in this section we argue that the respect of privacy principles is actually a core component of the trust underlying the relationship between users and service providers in a mature market fueled by big data. None of the parties benefits from interruption of this trust: if users feel that their personal data are not adequately protected, they will finally move towards solutions that "correct" this problem (although this might take some time). An example that greatly demonstrates this issue is that of behavioural advertising based on cookies. Since advertising companies and networks failed to adopt

---

[16] This is in fact already happening, see for example the latest Symantec report on the State of Privacy, https://www.symantec.com/content/en/us/about/presskits/b-state-of-privacy-report-2015.pdf
[17] As it happens in any commoditized market, see for instance John Quelch, *When Your Product Becomes a Commodity*, Harvard Business School 2007, http://hbswk.hbs.edu/item/5830.html

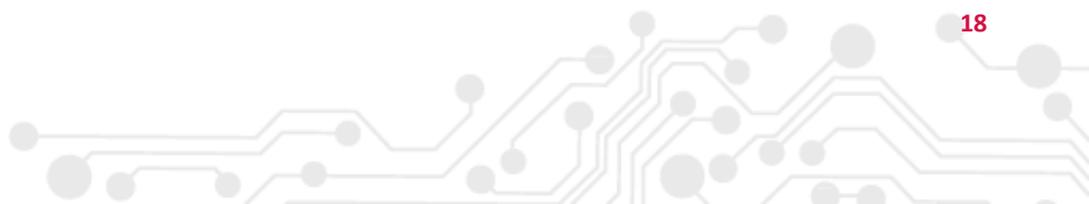

Privacy by design in big data
FINAL | 1.0 | Public | December 2015appropriate user-friendly consent mechanisms [19], the users slowly moved towards the wide adoption of ad blocking software [20] (which actually have much greater impact on the advertising industry).

*Privacy challenges should be, thus, seen as opportunities that, if appropriately handled, can build trust in the big data ecosystem for the benefit of both users and big data industry.*

For example, as already mentioned, the potential reuse of data is one of the key concerns in big data. How can individuals know if their data, collected for a specific purpose by a data controller, is being reused and for what purpose? According to the privacy principle of purpose limitation, data controllers who use personal data for analytics must ensure that the analysis is not incompatible with the original purpose for collecting the data. It is in fact essential that any arbitrariness or ambiguity is removed when defining the scope of the processing. In this exercise, aspects as the timescale of the processing and the impact on individuals should be considered[18]. Therefore, *transparency* to individuals and the provision of *meaningful tools to express their choice* are the only ways to control and perform data reuse. And also the only ways to gain users' trust on the processing. Any undesired reuse of personal data would be detrimental for the level of trust on the industry and at the end it would turn against the industry itself.

Another example is data inference and re-identification, which is becoming a real option today, with methodologies, and even software tools and computing power at the disposal of any attacker. How can an individual know that he/she will not be among the "special" cases of such an attack? And which service provider can he/she trust in this respect? Wide scale data breaches, the Ashley Madison case[19] being among the most recent ones, show that the disclosure of personal data can be disastrous not only for the individuals but also for the data controllers. As it is already emerging from measurements, big data will likely follow a known phenomenology for security incidents, the so called "black swan effect", oscillating between negligible cases (giving the impression that there is no new real risk around) and "big ones", where the impact to individuals will really matter[20]. Therefore, searching for correlations, which is supposed to be one of the major appeals of big data analytics, cannot be stretched above a threshold where it becomes more a privacy risk factor than a benefit (for individuals and data controllers). Embedding privacy enhancing technologies, such as *proper anonymization* and *encryption techniques*, in the analytics design is an integral part of avoiding personal data breaches and rebuilding the trust between users and service providers.

An additional example, especially relevant to social networks analytics, is the case of "context collapse" [21], namely the persistence of content, identifying and characterizing an individual, beyond the temporal moment of its creation and widely available and searchable by anyone. Against this potential risk, many studies have correctly pointed out the multi-faceted nature of our identities, which fail to be constrained in such schematic representations [22]. Still, how can the individual avoid situations where a piece of information follows him/her forever, even if this information is no longer valid? And how can he/she trust online services to publish his/her opinions or ideas under the threat that his/her own words may turn against him/her? Thus, an education towards compliant and lawful data processing is needed, keeping in mind that the interpretation coming out from big data analytics is "one possible" (probably the most likely) interpretation, but not the "only possible" interpretation, and other explanations of phenomena can exist. Transparency on the use of personal data is again central in this approach. Also the provision of *practical*

---

[18] See examples and guidance in Article 29 Working Party Opinion 3/2013 on purpose limitation [232].
[19] See: https://en.wikipedia.org/wiki/Ashley_Madison_data_breach
[20] See: S Wheatley, T Maillart, D Sornette, The Extreme Risk of Personal Data Breaches & The Erosion of Privacy arXiv preprint arXiv:1505.07684 (May 2015), http://arxiv.org/pdf/1505.07684.pdf. Also, Nassim Nicholas Taleb, The Black Swan: The Impact of the Highly Improbable, London: Penguin (2007).

*ways to users to access, modify and/or delete their data*, so as to make them feel in control. Control brings trust and again this is for the benefit of all involved parties.

These aspects come along with the relevance of automated decision processes in big data analytics. Many harmful consequences on data subjects may in fact occur with persistent, even if not intentional, discriminatory features [23], if a decision process is fully automated. This can even lead to the already observed "filter bubbles" effect [24], according to which data subjects will only be exposed to content which confirms their own attitudes and values, without any door open to serendipity and casual discovery[21]. Of course, data subjects will again likely react to these extreme scenarios, but also the industry may be worried by such unintended consequences of profiling. If one looks, for instance, at the current trends of the advertising revenues in the most advanced markets[22], it is not always clear whether the economic value of targeted advertising (based on profiles) really outweighs traditional contextual advertising (based on serendipitous choices). The quality of the profiles and the quality of personal data on which they are built, again, seem to matter for the prosperity of the industry, yet another relevant privacy principle.

Privacy is therefore crucial to protect the value of the data that will be used in big data analytics. Finding a balance between innovation and data protection without running the risks of over-reaction is extremely difficult. Big data analytics is an enormous opportunity for modern society and *privacy may be the touch of wisdom to discover the hidden consequences of a new technology, where others just see the dangers or only perceive the immediate rewards*. That is why privacy seems to be more than necessary today to provide guidance towards a data subject-centric development of big data.

### 2.4.3 Technology for big data privacy

In Europe there is already a strong legal framework for the protection of personal data which will soon be reformed to include new rights and obligations greatly applicable to the big data era. Recent ECJ decisions have addressed important legal matters to this end, especially regarding the establishment of the controller, the deletion of online data and the transfer of data outside the EU [25]. Still, regulation, no matter how strict or detailed it is, shows its limits in big data, as in the case of cloud computing and beyond this. Audit and enforcement are quite difficult in scenarios with diverse controllers and in territories where the national DPAs do not have inspection powers. The implementation of legal obligations is not always practical or easy to achieve in the complex technological landscape of big data. Therefore, although regulation is of utmost importance in setting the rules, it should not be considered as the only resort for the protection of personal data.

In this chapter we argued that big data is about scale. And scale is enabled by technology. *So, with respect to the underlying legal framework, technology (for big data) should be addressed by technology (for privacy)*. We cannot deal with new processing operations using old solutions. Therefore, measures for the protection of personal data need to grow together with big data advances. This is in fact the vehicle to implement the legal obligations and support enforcement. Obviously, we are not there yet. Still, we believe that there are already privacy preserving technologies and tools that can be applied in the era of big data and/or should be further explored for future use, following creative thinking and development. The concept of privacy and data protection by design is central in this approach and this will be our focus in the next chapter of this document.

---

[21] Interestingly, the archetype of big data analytics, the Google search Pagerank algorithm, seems to take into account serendipity, by allowing that user may escape at any time with some probability from the hyperlink structure of the web, to follow their own randomly selected route. See for instance A. Langville and C. Meyer. *A survey of eigenvector methods for web information retrieval*. SIAM Review, 47(1):135–161, 2005.

[22] Warc International Ad Forecast 2014/15, http://www.warc.com

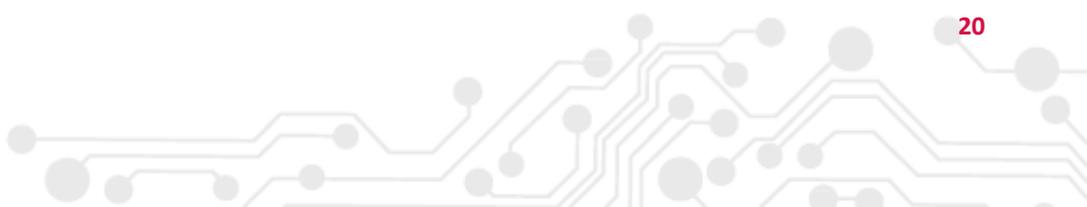

# 3. Privacy by design in big data

Finding the appropriate mechanisms to implement privacy principles in the big data environment is the most effective way to prevent a clash between privacy and big data which would have no winners. To this end, the concept of privacy and data protection by design is fundamental, as a mechanism to address the privacy risks from the very beginning of the processing and apply the necessary privacy preserving solutions in the different stages of the big data value chain. In this way, privacy by design can be a tool for empowering the individuals in the big data era, as well as supporting the data controllers' liability and trust.

This, however, is not always an easy task and many relevant questions arise. How can privacy by design be applied in big data processing and how can it practically support the implementation of the privacy principles and the underlying legal obligations? What types of privacy enhancing technologies are currently applicable to big data? Is privacy by design in big data a concept for today or tomorrow?

Based on ENISA's 2014 report on "Privacy and Data Protection by Design" [6], this chapter aims at addressing the above questions by presenting an overview of privacy by design in the big data value chain.

## 3.1 Privacy by design: concept and design strategies

Privacy by design was first widely presented by Ann Cavoukian [26], and pertained the notion of embedding privacy measures and privacy enhancing technologies (PETs) directly into the design of information technologies and systems. Nowadays, it is regarded as a multifaceted concept: in legal documents on one hand, it is generally described in very broad terms as a general principle; by computer scientists and engineers on the other hand it is often equated with the use of specific privacy enhancing technologies. However, privacy by design is neither a collection of mere general principles nor can it be reduced to the implementation of PETs. In fact, it is a process involving various technological and organizational components, which implement privacy and data protection principles.

In its 2014 report [6], ENISA explored the concept of privacy by design following an engineering approach. Among other things, the report, using relevant work in the field, presented eight privacy by design strategies, both data oriented and process oriented, aimed at preserving certain privacy goals. Table 1 provides a brief overview of the proposed design strategies.

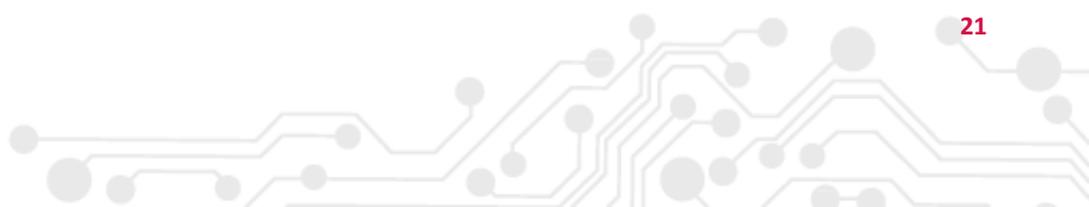

|   | **PRIVACY BY DESIGN STRATEGY** | **DESCRIPTION** |
|---|---|---|
| 1 | Minimize | The amount of personal data should be restricted to the minimal amount possible (data minimization). |
| 2 | Hide | Personal data and their interrelations should be hidden from plain view. |
| 3 | Separate | Personal data should be processed in a distributed fashion, in separate compartments whenever possible. |
| 4 | Aggregate | Personal data should be processed at the highest level of aggregation and with the least possible detail in which it is (still) useful. |
| 5 | Inform | Data subjects should be adequately informed whenever processed (transparency). |
| 6 | Control | Data subjects should be provided agency over the processing of their personal data. |
| 7 | Enforce | A privacy policy compatible with legal requirements should be in place and should be enforced. |
| 8 | Demonstrate | Data controllers must be able to demonstrate compliance with privacy policy into force and any applicable legal requirements. |

**Table 1: Privacy by design strategies [6]**

Moreover, following the privacy by design strategies, the report focused on a number of privacy enhancing technologies that can be used for implementing the strategies. Such technologies include authentication, attribute based credentials, secure private communications, communications anonymity and pseudonymity, privacy in databases, statistical disclosure control, privacy preserving data mining, private information retrieval, storage privacy, privacy preserving computations, transparency enhancing computations and intervenability enhancing technologies. The overall scope of the work was to bridge the gap between the legal framework and the available technologies, so as to allow for proper implementation of relevant solutions.

## 3.2 Privacy by design in the era of big data

Since privacy by design is about building privacy features at the very beginning of the processing, it can be a very useful concept and process for big data, allowing for early implementation of relevant controls that are protecting the individuals' personal data by default. Still, in the case of big data, due to the size and diversity of data being processed, even in near real time, several challenges are introduced.

First of all, in order to identify trends, detect patterns and reach valuable findings, the available data sets in big data should be as rich as possible. On the other hand, data minimization and limits in data retention seem to be an integral part of a privacy by design approach (the MINIMIZE design strategy in Table 1). Therefore, it can be argued that data minimization contravenes big data where large volumes of data are collected and stored before being used. Indeed, some of the most successful examples of big data come from digital breadcrumbs left behind by individual's use of technologies and later repurposed for analysis [27]. It is, thus, a major challenge in big data to minimize data collection, allowing at the same time for a useful rich content that can be used for analytics.

Moreover, the merging of information from different sources is also an essential part of big data analytics, going at first sight against the idea of distributed processing of personal data (the SEPARATE design strategy).

On top of this, the possibility to infer information and de-identify the individuals by the correlated data sets contradicts the very idea of hiding the data (the HIDE design strategy) [28]. Use of personal data in big data analytics is not always performed with the appropriate level of aggregation (the AGGREGATE design strategy).

In addition, the perpetual loop of repurposing and making use of already processed or inherent data, also affects the existing notice and consent models, as the processing sequence appears opaque to data subjects (against the privacy by design strategies INFORM and CONTROL). It can also be argued that even the data controllers themselves are not quite confident of the terms that their initial data set has been shaped, due to the chain of supply and processing of data by different parties (in contrast to the privacy by design strategies of ENFORCE and DEMONSTRATE).

So, is the privacy by design concept viable in the age of big data? Despite some arguing that this is not totally possible[23], our approach in this report is to actually revert the question: Can big data processing adopt to and benefit from the privacy by design approach and how can this be done in practice? The challenges are indeed present but, as discussed in Chapter 2, nobody will benefit from a clash between big data and privacy. If privacy is seen as a core value of big data, privacy by design can be a powerful tool. The MINIMIZE strategy would then provide for better and more useful data collection. The HIDE, AGRREGATE and SEPARATE strategies could allow for using personal data in analytics without affecting the individuals' private sphere. The INFORM strategy would support better mechanisms for users' information and transparency and the CONTROL strategy would support new practical ways for expressing consent and privacy preferences. Last, the ENFORCE and DEMONSTRATE strategies would support the data controllers in applying their privacy policies, in line with the principle of accountability.

Against this background, in the following we try to demonstrate how the privacy by design strategies can fit in the big data analytics landscape in a practical way.

## 3.3 Design strategies in the big data analytics value chain

When looking into the value chain of big data analytics, it is important to take into account the purpose of each of the phases and the standpoints of all parties involved there in (data subject, different data controllers and processors, third parties). In this way, it is possible to extract the specific privacy requirements and the relevant implementation measures per phase. Still, it is important to note, that apart from the needs of each particular phase, it is essential to implement a coherent approach to big data privacy protection, taking into account the complete lifecycle of the analytics. Therefore, it is not just about one technology or another, but rather about the synthesis of many different technologies adequately addressing all the needs of the different data processing nodes.

**Data acquisition/collection**

MINIMIZE: One very important privacy principle directly related to the big data collection phase is that of data minimization. Each data controller who is collecting the data needs to precisely define what personal data are actually needed (and what is not needed) for the purpose of the processing, including also the relevant data retention periods. Specific processes should be in place to exclude unnecessary personal data from collection/transfer, reduce data fields and provide for automated deletion mechanisms. Privacy Impact Assessments can be valuable tools for data controllers, in order to define the exact data processing needs limiting data to what is absolutely necessary for the certain purpose.

---

[23] See for instance [228] and [149] about the limits of data minimization, deletion and de-identification in big data.

AGGREGATE: Another aspect, which may come even as a result of the PIA, is when aggregated information is used instead of personal data. Indeed, in certain cases, such as in statistical analysis from distributed sources, the personal data might not even need to be collected in the first place and the collection of anonymised information might be sufficient. Local anonymization (or anonymization at source) is the most prominent solution, which could allow the individual (or a controller processing data for the individual) to remove all personal information before releasing the data for analytics.

HIDE: In many cases information about the individual may be collected without him/her even being aware (e.g. relating to his/her web searches and overall online behaviour). Privacy enhancing technologies that can support internet and mobile users' privacy are available today, including anti-tracking, encryption, identity masking and secure file sharing tools.

NOTICE: The individuals need to be adequately informed about the collection of their personal data for big data analytics. To this end, appropriate information notices and other transparency mechanisms should be in place. Such tools should be available to the individuals throughout the big data processing (and not only the collection). Still the point of collection is the most important point for the individual in order to make an informed decision about the use of his/her personal data.

CONTROL: Again the collection phase is the phase where the consent of the user needs to be obtained (if this is the legal basis for the processing). Practical and usable implementations of opt in mechanisms are crucial to this regard. Moreover, opt out tools should be offered to the individuals at any point of the processing. Other mechanisms providing user control, such as sticky policies and personal data stores, are important measures to explore.

**Data analysis and curation**

AGGREGATE: One of the most prominent techniques in the context of big data analysis is that of anonymization. Different privacy models and anonymization methods are in place to preserve data inference, for instance in statistical disclosure control and privacy preserving data mining techniques, including association rule mining, classification and clustering. K-anonymity and differential privacy are the two main families of privacy models with different types of implementations. Several cases of the application of such techniques exist today for big data, e.g. in the areas of social networks, health and location based services.

HIDE: Another very important technique in privacy preserving analysis is encryption, especially in the context of performing searches and other computations over encrypted data so as to protect individuals' privacy. Searchable encryption, homomorphic encryption and secure multiparty computations are promising technologies in this field with a lot of interest for the research community.

**Data storage**

HIDE: Security measures such as granular access control and authentication are essential for protecting personal data in databases. Technologies such as Attribute Based Access Control can be much more scalable in big data, offering fine grained access control policies. Encryption is also core in protecting data at rest.

SEPARATE: Privacy preserving analytics in distributed systems are also important for the protection of personal data as they provide for computations across different databases without the need for central warehouses. Access control measures and encryption techniques can again support this type of solutions.

**Data usage**

AGGREGATE: Privacy preserving data publishing and retrieval are usually based on anonymization in order to prevent inference of personal data. Issues related to data provenance in the course of decision making (based on big data) is another topic of interest, especially regarding the credibility and the level of aggregation of metadata (so as to avoid identification of individuals).

On top of the design strategies and controls mentioned above, it is important to point out that the data controllers and processors must first of all take into account the underlying legal obligations, in particular relating to the privacy principles and legal basis of the processing, as described in Chapter 2. To this end, the privacy by design principles ENFORCE and DEMONSTRATE are applicable to *all* phases of the big data value chain. Automated policy definition, enforcement and compliance tools can be useful in this task, supporting liability and accountability.

Following the aforementioned description, Table 2 provides an overview of the privacy by design strategies (from Table 1) and their possible implementation measures in each of the phases of the big data value chain.

| | BIG DATA VALUE CHAIN | KEY PRIVACY BY DESIGN STRATEGY | IMPLEMENTATION |
|---|---|---|---|
| 1 | Data acquisition/collection | MINIMIZE | Define what data are needed before collection, select before collect (reduce data fields, define relevant controls, delete unwanted information, etc), Privacy Impact Assessments. |
| | | AGGREGATE | Local anonymization (at source). |
| | | HIDE | Privacy enhancing end-user tools, e.g. anti-tracking tools, encryption tools, identity masking tools, secure file sharing, etc. |
| | | INFORM | Provide appropriate notice to individuals – Transparency mechanisms. |
| | | CONTROL | Appropriate mechanisms for expressing consent. Opt-out mechanisms. Mechanisms for expressing privacy preferences, sticky policies, personal data stores. |
| 2 | Data analysis & data curation | AGGREGATE | Anonymization techniques (k-anonymity family, differential privacy). |
| | | HIDE | Searchable encryption, privacy preserving computations. |
| 3 | Data storage | HIDE | Encryption of data at rest. Authentication and access control mechanisms. Other measures for secure data storage. |
| | | SEPARATE | Distributed/ de-centralised storage and analytics facilities. |
| 4 | Data use | AGGREGATE | Anoymisation techniques. Data quality, data provenance. |
| 5 | All phases | ENFORCE/ DEMONSTRATE | Automated policy definition, enforcement, accountability and compliance tools. |

Table 2: Privacy by design strategies in the big data value chain

It should be mentioned that the presentation of the design strategies is not explicit and it mainly aims at visualising the core principles that should be in place with regard to privacy and data protection upon the different phases of big data processing.

Against this background, in the next Chapter we examine some selected privacy enhancing technologies and solutions that we find particularly interesting in the case of big data.

# 4. Privacy Enhancing Technologies in big data

Following the privacy by design strategies described in Chapter 3, in this Chapter we provide an overview of key identified technologies that could be further applied and/or explored in the particular case of big data. Having said that, it is important to note that most of these technologies are available today and have been used in the "traditional" processing of personal data [6]. But again it is all about the scale: in the context of this report we aim to explain the specificities and adoption of certain technologies in big analytics, taking account the volume, velocity, variety and veracity of the new data landscape.

In particular, we first present anonymization, which has been so far the "traditional" approach towards data analytics, facing some new challenges in the era of big data. Then we explore the developments in cryptography and more specifically encrypted search, which can allow for privacy preserving analytics without disclosing personal data. On top of these approaches, we discuss privacy by security, i.e. ensuring an overall security framework for the protection of personal data, especially regarding access control policies and their enforcement. Transparency and control mechanisms are also central in big data, in order to offer information and choice to the individuals. To this end, we examine notice, consent and other mechanisms relying on users' privacy preferences and taking into account relevant usability issues.

## 4.1 Anonymization in big data (and beyond)

Anonymization refers to the process of modifying personal data in such a way that individuals cannot be re-identified and no information about them can be learned[24]. It is applied in several cases of data analysis, especially in the context of Statistical Disclosure Control (SDC [29]).

Perfect anonymization is difficult in practice without compromising the utility of the data set. In big data this problem increases due to the amount and variety of data. On one side, low level anonymization (e.g. mere de-identification by just suppressing direct identifiers) is usually not enough to ensure non-identifiability [13][30]. On the other side, too strong anonymization may prevent linking data on the same individual (or on similar individuals) that come from different sources and, thus, thwart many of the potential benefits of big data.

In this section, we review anonymization trade-offs and techniques. Most of what we cover refers to any kind of data, not specifically big data: it is about techniques to prevent re-identification (singling out), attribute disclosure and linking the records corresponding to the same individual [17]. Still, we treat the following aspects of anonymization that are specific to big data:

- **Controlled linkability**: Even if preventing linking records is one usual goal of anonymization [17], methods that fulfil the other two goals (preventing re-identification and attribute disclosure) while allowing some linkability are of interest in the case of big data. Indeed, big data anonymization should be compatible with linking data from several (anonymized) sources.

- **Composability:** This property refers to the anonymization privacy models: a privacy model is composable if its privacy guarantees hold for a data set constructed by linking together several data sets for each of which the privacy guarantee of the model holds. Of course, composability is very important for big data, where data sets are formed by merging data from several sources.

---

[24] It is important to note that the concept of anonymization is stronger than that of de-identification, which refers only to the removal of possible identifiers from the data set.

- **Anonymization of dynamic/streaming data**: Most of the anonymization literature refers to static data sets. However, in big data it is very frequent to encounter continuous data streams, for example, the readings of sensors. We examine utility and disclosure risk control in the case of streams of data.

- **Computability for large data volumes**: In big data, even static data sets may be challenging to anonymize due to their sheer volume. Hence, computational efficiency may be a critical issue when choosing a privacy model or an anonymization method.

- **Decentralized anonymization**: Under this paradigm, the data subject anonymizes his/her data at the source, using his/her personal computing device, before releasing those data to the data controller. This reduces the need for trust by data subjects to the controller. We review two forms of decentralized anonymization: local anonymization and collaborative co-utile anonymization.

After a general presentation on anonymization techniques, the next paragraphs present the big data specific topics, together with relevant open issues and trends in the field.

### 4.1.1 Utility and privacy

Given that anonymization methods modify the original data to prevent disclosure of personal information, a tension clearly arises between utility and privacy. The challenge is to protect privacy with minimum loss of accuracy: ideally, data users should run their analyses on the anonymized data without losing accuracy with respect to the results of those analyses when run on the original data.

The question then becomes: how to measure the utility of an anonymized released data set?

Data-use specific and generic utility measures can be used to address this question (see Annex 2 for more details). In the case of big data, the same divergence measures can be applied, replacing standard algorithms used for analysis with specific ones used for clustering and classification of big data (including streaming data). Nevertheless, computational difficulties arise when measuring utility and information loss, due to the computational problems of applying the analysis, the variability/soundness of the result (e.g. the stability of the clusters found when clustering big data), and also the computation of the divergence.

For specific types of data with a large volume, some particular data uses exist and this permits the definition of new utility measures. See, for example [31] for analysis and measures on graphs to study social networks. These measures can be used to evaluate information loss and data utility of protected graphs.

#### 4.1.1.1 A specific utility measure for big data: linkability

Without being specific to any particular analysis, linkability is key to obtain information from the fusion of data collected by several sources. In big data, information about an individual is often gathered from several independent sources. Hence, the ability to link records that belong to the same (or of the same type/similar) individual is central in big data creation.

With privacy protection in mind, the data collected by a given source should be anonymized at the source before being made available. However, independent anonymization by each source may limit data fusion capabilities, thereby severely restricting the range of analyses that can be performed on the data and, consequently, the knowledge that can be generated from it. The amount of linkability compatible with an anonymization technique or with an anonymization privacy model determines whether and how an analyst can link data independently anonymized (under that technique/model) that correspond to the same individual. Of course, while linkability is desirable from the utility point of view, it is also a privacy threat: the accuracy of linkages should be significantly less in anonymized data sets than in original data sets.

For more details on how to measure linkability with different anonymization privacy models or anonymization techniques, see [32].

#### 4.1.1.2 Utility-first versus privacy-first approach

Broadly speaking, there are two approaches for anonymization to deal with the tension between utility and privacy:

- **Utility-first anonymization:** an anonymization method with a heuristic parameter choice and with suitable utility preservation properties is run on the microdata set and, after that, the risk of disclosure is measured. For instance, the risk of re-identification can be estimated empirically by attempting record linkage between the original and the anonymized data sets [33][34][35]. If the extant risk is deemed too high, the anonymization method must be re-run with more privacy-stringent parameters and probably with more utility sacrifice.

- **Privacy-first anonymization:** In this case, a privacy model is enforced with a parameter that guarantees an upper bound on the re-identification disclosure risk and perhaps also on the attribute disclosure risk. Model enforcement is achieved by using a model-specific anonymization method with parameters that derive from the model parameters. Well-known privacy models include k-anonymity and its extensions, as well as ε-differential privacy (we review them below). If the utility of the resulting anonymized data is too low, then the privacy model in use should either be enforced with an alternative anonymization method that is less utility-damaging, or a less strict privacy parameter should be chosen, or even a different anonymization privacy model should be resorted to.

In practice, most data releasers today (e.g. national statistical offices) tend to adopt the utility-first approach, because delivering useful data is their "raison d'être". The privacy-first approach, based on anonymization privacy models, has initially been presented by academic computer scientists working in the area of privacy. However, attempts at using privacy models in actual data releases have been made (for example, differential privacy in [36] and *k*-anonymity in [37]), although with parameter choices relaxing privacy in order for reasonable utility to be attainable. The privacy-first approach has also been adopted by the Working Party 29 [17] on their opinion on anonymization of personal data.

### 4.1.2 Attack models and disclosure risk

In the context of anonymization, privacy can be compromised by means of two types of disclosure: identity disclosure and attribute disclosure. Most attacks and privacy paradigms can be categorized as focusing on one or the other type.

- **Identity disclosure**: An intruder is able to link the data in a published data set with a particular individual (also known as entity disclosure).

- **Attribute disclosure**: Intruders improve their knowledge on the value of an attribute of an individual. Attribute disclosure can also be considered in the case of an intruder who finds out that an individual's data are included in a database. Attribute disclosure was already defined in [38] by Dalenius.

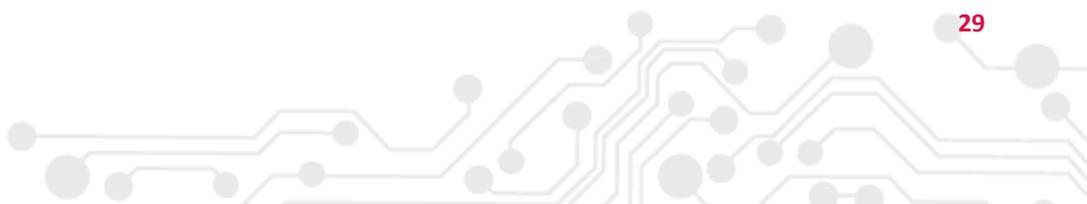

Although it is more common that identity disclosure implies attribute disclosure, identity disclosure and attribute disclosure are two different notions. Identity disclosure does not require attribute disclosure to take place when all the available information is used in the linkage process. Attribute disclosure can be observed without identity disclosure when individuals share the same value on a given attribute (see *l-diversity* below).

The study of disclosure risk is a controversial topic, on both notions identity and attribute disclosure. The main challenge that attribute disclosure poses concerns statistics and data mining analysts, who expect that an increase on their knowledge occurs after a data release. Data release should increase knowledge not only for populations, but also for particular individuals. For example a model inferred from a released data set has enough detail so that it permits to increase our accuracy on the values of certain attributes for given individuals [39]. The practical impact of identity disclosure is yet another topic for discussion. There are some arguments that the risk of identity disclosure is exaggerated and that disclosure risk should not prevent data release [40][41]. Still, as mentioned in Chapter 2, a single data breach is enough to bring the pieces of information together and lead to the re-identification of one or more individuals.

In any case, an accurate analysis of disclosure risk is necessary prior to any release of information. See Annex 2 for more information on measures of disclosure risk.

### 4.1.3 Anonymization privacy models

Most privacy models can be classified in two main families. A first family includes *k*-anonymity [42] and its extensions taking care of attribute disclosure, like p-sensitive *k*-anonymity [43], *l*-diversity [44], *t*-closeness [45], *(n,t)*-closeness [46], and others. The second family is built around the notion of *ε*-differential privacy [47], along with some variants like crowd-blending privacy [48] or BlowFish [ [49]. We briefly review those models in the next paragraphs (see the above references for more details); some of these privacy models are also reviewed in [17]. Re-identification based models, which can be considered a third family (and are closely related to k-anonymity), are also reviewed in Annex A2.

#### 4.1.3.1 k-Anonymity and its extensions

The *k*-anonymity family of models is based on a classification of the attributes of a data set into several non-disjoint types:

- Identifiers. These are attributes in the original data set that unambiguously identify the data subject to whom a record corresponds. Examples are passport number, social security number, full name, etc. Identifiers are removed as a precondition towards obtaining an anonymized data set.

- Quasi-identifiers (also known as key attributes). These are attributes in the original data set that, in combination, can be linked with external information to re-identify (some of) the subjects to whom (some of) the records in the original data set refer. Examples are job, age, city of residence, etc. Unlike identifiers, quasi-identifiers cannot be removed as part of the anonymization process, because any attribute is potentially a quasi-identifier.

- Confidential attributes. These are attributes that contain sensitive information on the data subject. Examples are salary, religion, health condition, etc.

*k-Anonymity*. A data set is said to satisfy *k*-anonymity for *k>1* if, for each combination of quasi-identifier attribute values, at least *k* records exist in the data set sharing that combination.

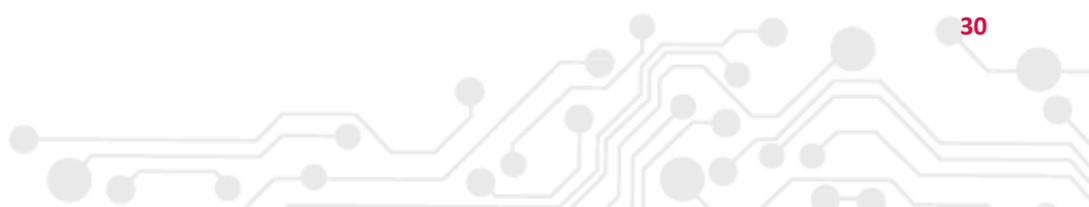

*k*-Anonymity is able to prevent identity disclosure, that is, a record in the *k*-anonymized data set cannot be mapped back to the corresponding record in the original data set. Indeed, the *k*-anonymous version of an original record is indistinguishable within a group of *k* records sharing quasi-identifier values. However, in general, *k*-anonymity may fail to protect against attribute disclosure; this is the case if the value of a confidential attribute is the same or very similar in all *k* records sharing the same combination of quasi-identifier values. We next review several extensions of *k*-anonymity whose aim is to prevent attribute disclosure.

*p-Sensitive k-anonymity*. A data set is said to satisfy *p*-sensitive *k*-anonymity for *k>1* and $p \leq k$ if it satisfies *k*-anonymity and, for each group of records with the same combination of quasi-identifier attribute values, the number of distinct values for each confidential attribute within the group is at least *p*.

*l-Diversity*. A data set is said to satisfy *l*-diversity if, for each group of records sharing a combination of quasi-identifier attribute values, there are at least *l* "well-represented" values for each confidential attribute. Several definitions of "well-represented" are suggested in [44]: a) the *l* values are merely distinct (in this case *l*-diversity is equivalent to *l*-sensitive *k*-anonymity); b) Shannon's entropy of the confidential attribute values within each group is at least $log_2 l$; c) recursive *l*-diversity, which requires that the most frequent values do not appear too frequently and the least frequent values do not appear too rarely.

*p*-Sensitive *k*-anonymity and *l*-diversity do not take into account the semantics of the confidential attribute values and are therefore vulnerable to skewness and similarity attacks. In the former, there are *l* diverse values, but a lot of these are skewed (e.g. high salaries in an attribute "salary" or serious diseases in an attribute "diagnosis"). In the latter, there are *l* diverse values, but a lot of them are very similar from the semantics point of view (e.g. a lot of serious cancers in an attribute "diagnosis"). See [50] for more details. Such attacks are better countered by *t*-closeness, defined as follows.

*t-Closeness*. A data set is said to satisfy *t*-closeness if, for each group of records sharing a combination of quasi-identifier attribute values, the distance between the distribution of the confidential attribute in the group and the distribution of the attribute in the whole data set is no more than a threshold *t*.

*(n,t)-Closeness* is a relaxation of *t*-closeness. In this case, for each group of records sharing a combination of quasi-identifier attribute values, the distance between the distribution of the confidential attribute in the group and the distribution in a superset of the group with at least *n* records should be no more than a threshold *t*.

### 4.1.3.2 Differential privacy and related models

Differential privacy is a privacy model that seeks to limit the impact of any individual subject's contribution on the outcome of the analysis. Its primary setting was to anonymize the answers to interactive queries submitted to a database, rather than anonymizing data sets. For this reason it is of particular interest to big data.

*ε-Differential privacy*. A randomized function κ (a function that returns the query answer plus some noise) satisfies *ε*-differential privacy if, for all data sets $D_1$ and $D_2$ that differ in one record (neighbor data sets), and all $S \subset Range(\kappa)$, it holds that

$$\Pr(\kappa(D_1) \in S) \leq \exp(\varepsilon) \times \Pr(\kappa(D_2) \in S))).$$

Several proposals have been made to generate differentially private data sets [51][52][36][53][54]. They follow two main approaches: (i) create a synthetic (simulated) data set from an *ε*-differentially private for the data (usually from a differentially private histogram), or (ii) add noise to mask the values of the original records (probably in combination with some prior aggregation function to reduce the amount of required noise).

A downside of $\varepsilon$-differential privacy is that it provides a strong privacy guarantee at the expense of a severe utility loss. Indeed, if the presence of any original record needs to be unnoticeable within $\exp(\varepsilon)$ in the anonymized data set (which is basically what the differential privacy requirement says), it is hard to preserve any significant utility unless $\varepsilon$ is large (often much larger than 1), in which case the privacy guarantee is no longer that strong. See [55] for more details on utility-related criticisms to differential privacy.

*Crowd-blending privacy* can be viewed as a relaxation of differential privacy that is partly inspired on *k*-anonymity, with a view to improving utility. A data set satisfies *k*-crowd blending privacy if each individual record *i* in the data set "blends" with *k* other records *j* in the data set, in the sense that the output of the randomized query function κ is "indistinguishable" if *i* is replaced by any of the records *j*. Thus, similar to the relaxation of *t*-closeness into *(n,t)*-closeness, differential privacy is relaxed into crowd-blending privacy by transforming a requirement involving the entire data set to a requirement involving only a group of records containing a specific record.

*Blowfish* can both be regarded as a relaxation and a generalization of differential privacy: it uses the same condition, but it changes the definition of neighboring data sets. Whereas in differential privacy neighboring data sets $D_1$ and $D_2$ are defined as those differing in any single record, in Blowfish one can use any definition of neighborhood. If, as a result, the number of neighbors is a strict subset of those in standard differential privacy, then we have a relaxation.

In [54] a way to improve data utility in differential privacy based on microaggregation-based multivariate *k*-anonymity was presented. Further, in [56] it was shown how differential privacy can be reached from *t*-closeness and vice versa. This does not mean that both models are equivalent, but it illustrates that they can provide similar privacy guarantees if the proper parameter choices are made.

**4.1.4 Anonymization privacy models and big data**

In [57] linkability, composability and computability are identified as requirements that a privacy model must satisfy in order to be useful for big data anonymization. The above properties are also evaluated for *k*-anonymity and differential privacy.

In particular, *k*-anonymity offers linkability at the group level (among groups of *k* records), but is not composable (linking two *k*-anonymous data sets is not guaranteed to yield a *k'*-anonymous data set for any *k'>1*). For example, if we have two k-anonymous data sets of patients from two different hospitals (e.g. including zip code, age range and disease of the patient), it would be possible to identify a certain individual in the set, for instance if someone has information that this individual visited both hospitals and his/her age and location are known. Overall k-anonymity cannot guarantee privacy if sensitive values in the data set lack diversity and some further information is known to the attacker [58].

On the other hand, $\varepsilon$-differential privacy is strongly composable: combining an $\varepsilon_1$-differentially private data set and an $\varepsilon_2$-differentially private data set yields an $\varepsilon_1+\varepsilon_2$-differentially private data set (hence, differential privacy is still satisfied, although with a less strict parameter). Regarding linkability, differentially private data sets are not linkable if obtained using noise addition, but they can be made linkable if obtained using partially synthetic data generation (see below).

For both privacy models, computational complexity is very dependent on the anonymization method used to satisfy them.

Generally speaking, k-anonymity and differential privacy are often perceived as competing in big data (and beyond big data). K-anonymity has in fact received many criticisms regarding its weaknesses and differential privacy has been presented as the solution to this problem [59]. Still, it is important to the note that there is a fundamental difference between the two approaches [60]. K-anonymity (and its variants) is focusing on anonymizing a data set before its release for further analysis (e.g. a hospital that wants to publish data in order to analyze side effects of certain medications). Differential privacy, on the other hand, is about running queries on the data, following a predefined type of analysis, in a way that the answers do not violate individuals' privacy. Although it has been argued that the differential privacy's query-based approach is superior to the "release and forget" approach of k-anonymity [61], its practical implementation (taking into account the utility-privacy trade-off) is not possible in every data analytics scenario. Thus, k-anonymity still has a dominant role, especially on data releases (i.e. when the query-based model is not applicable).

In conclusion, both approaches have different pros and cons and further research is required to maintain a good balance between utility and privacy for different application scenarios in big data.

### 4.1.5 Anonymization methods

There are two principles used in microdata anonymization, namely data masking and data synthesis:

- Masking generates a modified version X' of the original microdata set X, and it can be perturbative masking (X' is a perturbed version of the original microdata set X) or non-perturbative masking (X' is obtained from X by partial suppressions or reduction of detail, yet the data in X' are still true).

- Synthesis is about generating synthetic (i.e. artificial) data X' that preserve some preselected properties of the original data X.

The above methods are reviewed in more detail in Annex 2. See [29] for a more comprehensive survey; a review of a subset of the aforementioned methods can also be found in [17].

### 4.1.6 Some current weaknesses of anonymization

Currently, there are several shortcomings in anonymization methods, which need to be further considered:

- **Comparability**: Comparing anonymization methods for microdata in terms of utility and disclosure risk is made difficult by the diversity of the principles they rely upon.

- **Verifiability by data subjects**: Current anonymization does not favour the data subject's information self-determination. The data controller takes legal responsibility for the release and makes all choices (anonymization method, parameters, privacy and utility levels, etc.). Moreover, the data subject cannot verify whether his/her record is getting adequate protection.

- **Adversary's background**: In the utility-first approach and the privacy-first approach based on the k-anonymity family, restrictive assumptions need to be made on the adversary's background knowledge (namely, one assumes that the adversary can only link with external identified data sets through a subset of quasi-identifier attributes). In ε-differential privacy, no restrictive assumptions need to be made, but enough perturbation is needed to make presence/absence of any particular record unnoticeable in the anonymized data; as a consequence, data utility is substantially damaged. Annex 2 provides more details on this topic (on permutation paradigm and maximum-knowledge adversary).

- **Transparency to users**: The question here is how much detail shall/can be given to the user on the masking methods and parameters used to anonymize a data release. Clearly, the user derives more inferential utility from increased detail [62]. Yet, some methods may be vulnerable if too much detail on them is given. Further discussion on transparency can be found in Annex 2, including transparency-safe anonymization methods, as well as the permutation paradigm of anonymization and its verifiability and transparency implications.

### 4.1.7 Centralized vs decentralized anonymization for big data

#### 4.1.7.1 Pros and cons of centralized anonymization

The mainstream literature on statistical disclosure control (e.g. [29]) focuses on centralized anonymization, performed by a data controller who has access to the entire original data set. This centralized approach has some advantages:

- Individuals do not need to anonymize the data records they provide. The data controller, who has more computational resources and probably more expertise in anonymization, can be expected to adequately anonymize the entire data set.

- The data controller has a global view of the original data set and, thus, is in the best position to optimize the trade-off between data utility and extant disclosure risk.

Still, there are also some important downsides to centralized anonymization:

- The data controller must be trusted by all parties providing original data (because the controller has access to all original data). While this is not a problem in official statistics, where the data controller is a national statistical institute, it can be a major hurdle in a typical big data scenario, for instance when the data controller assembling several data sources is merely a private company (e.g. a data broker).

- Especially in the case of big data, anonymization can be too heavy a computational burden for a single controller.

- Many controllers are involved in a single big data processing scenario, thus, making the centralised approach unmanageable.

The above downsides have prompted the design of local anonymization approaches and, very recently, collaborative anonymization.

#### 4.1.7.2 Local anonymization

Local anonymization is an alternative disclosure limitation paradigm suitable for scenarios (including big data) where the individuals (data subjects) do not trust (or trust only partially) the data controller assembling the data. Each subject anonymizes his/her own data before handing them to the data controller. In comparison to centralized anonymization, local anonymization usually results in greater information loss (data utility loss). The reason is that each individual needs to protect his/her data without seeing the other individuals' data, which makes it difficult for him/her to find a good trade-off between the disclosure risk limitation achieved and the information loss incurred.

We briefly review some local anonymization techniques. Many standard SDC techniques can be applied locally, such as generalization, top/bottom coding and noise addition. Among the techniques specifically designed for local anonymization, the oldest one is probably randomized response [63]. In randomized response, the data subject flips a coin before answering a sensitive dichotomous question (like "Have you taken drugs this month?"); if the coin comes up tails, the data subject answers "yes", otherwise he/she answers the truth. This protects the privacy of individuals, because the data controller/collector cannot determine whether a particular "yes" is random or truthful, but he knows that the "no" answers are truthful, so it is possible to estimate the real proportion of "no" as twice the observed proportion (from which the real proportion of "yes" follows). FRAPP [64] can be seen as a generalization of randomized response. In FRAPP, the data subject reports his/her real value with some probability and, otherwise, returns a random value from a known distribution. In AROMA [65] each data subject hider his/her confidential data within a set of possible confidential values drawn from a known distribution; if all data subjects honestly follow the local anonymization procedure, the pooled anonymized data can satisfy differential privacy.

#### 4.1.7.3 Co-utile collaborative anonymization

As already mentioned, a problem with centralized anonymization is that, if a data subject does not trust the data controller to properly use and/or anonymize his/her data, he/she may decide to provide false data (hence causing a response bias) or no data at all (hence causing a non-response bias). Local anonymization is an alternative that is not free from problems either: for a data subject anonymizing his/her own record in isolation it is hard to determine the amount of masking that yields a good trade-off between disclosure risk and information loss. A natural tendency is for each individual to play it safe and overdo the masking just in case, which incurs more information loss than necessary.

A more realistic goal is to generate, in a collaborative and distributed manner, an anonymized data set that satisfies the following conditions: (i) it incurs no more information loss than the data set that would be obtained with the centralized approach for the same privacy level; and (ii) neither the data subjects nor the data controller gain more knowledge about the confidential attributes of any other specific data subject than the knowledge contained in the final anonymized data set. In [66] a protocol is given whereby a set of data subjects can collaboratively attain *k*-anonymity.

In fact, collaborative anonymization is feasible because it leverages the co-utility principle [67], that is, mutual utility. Note that the privacy protection obtained by a data subject positively impacts the privacy protection obtained by the rest of data subjects in the data set. Put in the negative: if one data subject from a set of *n* subjects is re-identified, the remaining *n-1* become easier to re-identify (as the crowd in which they are hiding gets thinner). Hence, data subjects are rationally interested in collaborating for anonymity.

### 4.1.8 Other specific challenges of anonymization in big data

In big data it is also important to distinguish between anonymization methods that deal with volumes of data (but with low or no variability), dynamic publishing (databases with variability and with the need of multiple publication of the same database), and streaming data (data arrives continuously and we need to process as it comes). The next paragraphs provide an overview of such techniques.

#### 4.1.8.1 Large volumes of data

Some masking methods have been defined for dealing with standard data files of large dimensions. In this case, special importance is given to efficiency. The authors of [68] give an overview of this problem, and, for example, [69] & [70] describe micro aggregation methods specially designed for large numerical data sets, [71] an efficient method for k-anonymity and [72] how to measure disclosure risk also for large data sets. In addition, there is a large set of methods that deal with particular data types that are typically big. This is the case of data privacy for online social networks, location privacy and text data. Literature for both social networks and location privacy follows the trends and research lines we have described so far.

For example, in the case of online social networks we have different anonymization methods. There are perturbative methods that introduce (by means of e.g. random noise, generalization and micro aggregation) noise to edges and vertices of the social network [73]. There are also methods that focus on achieving k-anonymity for graphs. For this purpose different definitions have been proposed [74] taking into account different types of attacks. The simplest type of attack is when the intruder knows the degree of a vertex. k-degree anonymity [75] focuses on k-anonymity for this type of attack, that is when the quasi-identifier is the degree. [76] considers that the intruder has information on the neighbors of a vertex and their relationships. We call k-neighborhood anonymous when the vertices of a graph is k-anonymous with respect to this information. [73] & [77] discuss k-anonymity with respect to structural queries. That is, they require that an anonymity set of at least k individuals is returned for a given but arbitrary query. Differential privacy has also been applied to online social networks (see e.g. [78]). In this context we have different types of differential privacy [79][80]. For example, node differential privacy and edge differential privacy. In node differential privacy the focus is on whether the presence or absence of a node changes the outcome of a query. In edge differential privacy, the focus is on the presence or absence of an edge.

Research also exists on the anonymization of textual data. Some techniques are based on the identification of proper names and locations, finding sensitive words and then marking or replacing them by more generic and/or semantically similar ones [81][82][83]. Other works focus on k-anonymized vectors of terms that can be later used for information retrieval-like applications [84]. For this purpose, syntactic and semantic (using ontologies and dictionaries) approaches have been considered.

Similar results have been obtained for anonymization of data in location-based services. See e.g. [85] on the location privacy through generalization, and [86] about differential privacy in the same context.

### 4.1.8.2 Dynamic data

Dynamic data is the case when a database changes with respect to time and data has to be published regularly. In such a scenario we can have disclosure of data if the different releases do not take into account that some information has already been published before (even if protected using e.g. k-anonymity). The problems of dynamic data publishing are discussed in [87] and algorithms for this purpose can be found also in [88] and [89]. Masking methods for dynamic publication of databases with textual documents are given in [90].

### 4.1.8.3 Streaming data

From a privacy perspective data streams are posing new challenges to all stakeholders involved in the big data analytics value chain. The first challenge is related to the incompleteness of information: due to the fact that data comes into the system in portions and potentially in an unstructured sequence, the evaluation of privacy preserving algorithms is difficult (as there is always some missing information to verify the results). The second challenge is the way that the model/patterns from the data are learned: in streams this happens online incrementally and is always updated, which makes standard anonymizing techniques not useful. Third challenge is the dynamic nature of data streams and the impact of temporal phenomena to fixed privacy rules. For example as the authors in [91] present "suppose winter comes, snow falls, and much less people commute by bike. By knowing that a person comes to work by bike and having a set of GPS traces, it may not be possible to identify this person uniquely in summer, when there are many cyclists, but possible in winter". Therefore static privacy rules cannot be used and new adaptive privacy preservation and enforcement mechanisms are required to successfully deal with data streams.

Most of the contributions today are in the area of differential privacy [92][93][94] and k-anonymity [95], [96][97], although there are also perturbative approaches [90][97]. Some of the k-anonymity and perturbative approaches are based on a sliding window, in which a masking method is applied. In this case, methods described in previous sections can be used.

### 4.1.9 Challenges and future research for anonymization in big data

Anonymization methods for static and structured data sets often have shortcomings, in particular related to comparability, verifiability, adversarial model, and transparency (see 4.1.6). Big data introduces additional challenges to these properties, as data formats expand to volatile and unstructured streams, such as data from meters, sensors or complex images from medical applications. The typical "release-and-forget" de-identification model [98] has shown its limits in big data and there are already many cases of high-dimensional data sets that have led to re-identification of individuals [13] [14] [28][99][100][101][102][103], e.g. in the context of mobile phones, internet of things, public transportation, genetics, credit card, or wearable devices.

Therefore, an issue that deserves further investigation is the validity of current anonymization privacy models and methods for big data. In particular, one needs to examine whether the privacy guarantees defined by current privacy models are compatible with data continuously and massively collected from multiple data sources. In other words, the question is whether such guarantees can be attained without a mass destruction of the collected data. In case the answer is negative, new anonymization privacy models and methods should be designed from scratch with big data requirements in mind.

Moreover, the difficulty of calculating the risk of inference significantly increases in big data. Intended to evaluate which data may be sensitive or cause harms, risk assessments are becoming increasingly hard to perform in big data, ultimately limiting their relevance. Indeed, when considering the risks associated with mobile phone, sensors, location, or spending data, adequate risk assessments will need to consider not only what is directly visible about an individual in the data, but also what an algorithm could uncover from the data, now or in the future. This issue was exemplified recently when mobile phone data was combined with machine learning algorithms to show that someone's sexual preference could be predicted from Facebook likes [104] or his or her personality from seemingly innocuous mobile phone data[25] [105]. Assessing the risk of inference in big data, thus, requires significant investments in specialized training data sets and fast-evolving machine learning techniques which need to be further developed and considered.

Taking into account the above mentioned challenges, it is important also to emphasise that anonymization cannot be regarded as a data protection guarantee per se. It is rather a primary component of safeguard in a picture which already incorporates other privacy safeguards for individuals, including lawful and transparent data collection and security of the overall processing. Moreover, anonymization cannot be the (sometimes easy) solution to all problems. In fact there are cases, for example when analytics are run in distributed systems in real time, that anonymization might not be applicable at all. In such cases the data are never released per se but accessed remotely by researchers, often through questions-and-answers systems. Different types of controls are in those cases needed so as to guarantee privacy and data protection in big data analytics.

---

[25] In this case, it was shown that an individual's personality traits – the big five: neuroticism, openness, conscientiousness, extraversion, agreeableness [229]– could be predicted up to 1.7 times better than random.

## 4.2 Encryption techniques in big data

Encryption is a fundamental security technique, which transforms data in a way that only authorised parties can read it, and a strong protection measure for personal data. Its role can be integral in big data, as long as it is performed using suitable encryption algorithms and key sizes, and the encryption keys are adequately secured[26]. Still, big data analytics need to allow for search and other computations over the stored data. How can this be achieved without decrypting the data (which would contradict the very idea of encryption)? In this section we cover selected developments in the field of operations in the encrypted domain; however, it should be mentioned that this selection is neither exhaustive nor is it based on a thorough assessment of the presented approaches. Moreover, most of these techniques are largely theoretical at the present time and it might take long before they become viable in real environments.

### 4.2.1 Database encryption

As already mentioned, encryption is an essential technical control, particularly in cloud computing big data environments. To this end, local encrypted storage [6] (full disk or file system level) is actually offered by many big data systems and solutions today. For instance in the Apache Hadoop ecosystem, a project called Rhino [106] offers transparent encryption to the Hadoop Distributed File System and to HBase tables of the system, something that is offered by Protegrity [107] as well. In NoSQL databases zNcrypt [108] offers application level data encryption while encryption keys are kept with the user [9]. Similarly, cipherDB offers this functionality to Microsoft Azure world [109].

Overall, it could be stated that symmetric encryption schemes based on AES are widely used in big data and cloud environments, due to their efficiency and security. However, there are some concerns related to secure and scalable key management, as well as to the possibility to perform certain functionalities without disclosing the secret key. Public key encryption schemes (e.g. RSA) are more demanding in terms of computing resources and they are mainly used in hybrid schemes for distributing secret keys (or for digital signatures). Hybrid schemes are predominant because they combine the advantages of public key encryption in scalability and key management with the speed and space advantages of symmetric encryption. They are often used in mobile environments with limited transmission of data and many users [110]. More information on the current application of encryption techniques in big data systems can be found in ENISA's 2015 reports on big data threat landscape and big data security[27].

On top of the "traditional" local encryption solutions, the state-of-the-art in big data storage requires encryption technologies going beyond the"encrypt all or nothing" model and offering finer grained data sharing policies. This is particularly important in order to allow different users to access different parts of the data, share information and perform the necessary analytics.

Attribute-Based-Encryption (ABE) is an emerging technique today, opening the road for sharing data among different user groups, while preserving users' privacy. In particular, ABE combines access control with public-key cryptography, in a way that the secret key used for the encryption and the ciphertext depend upon certain attributes (e.g. the individual's country, job or habit) [111]. In this way, the decryption of the ciphertext can be performed only if the presented set of attributes matches the attributes of the ciphertext.

---

[26] See ENISA's 2014 report on cryptographic algorithms and key sizes, providing guidelines for decision makers in designing and implementing cryptographic solutions for the protection of personal data, https://www.enisa.europa.eu/activities/identity-and-trust/library/deliverables/algorithms-key-size-and-parameters-report-2014
[27] See ENISA's web site for latest publications, https://www.enisa.europa.eu/

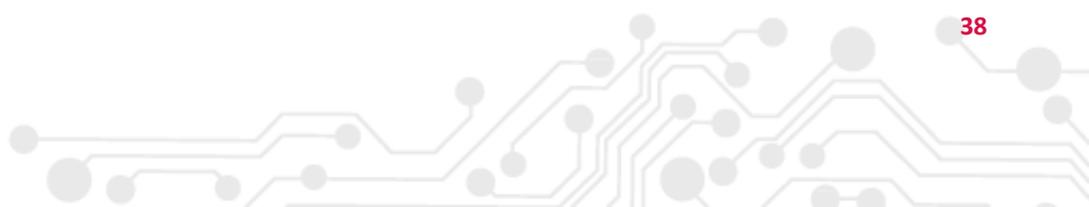

The simplest implementation of ABE is that of identity based encryption (IBE), where both the ciphertext and secret key are associated with identities (e.g. the user's email address) and decryption is possible exactly when the identities are equal [112].

Functional encryption [113] is an advancement of ABE where a user with certain attributes will have a key that will enable him/her to access a particular function of the encrypted data. Functional encryption is particularly interesting in cases where a set of cipher text can be seen by everyone, but only a specific portion of it can be decrypted and for certain processing that is mandated in the secret key itself [114]. For example, in the context of a medical study with encrypted patient data, it could be possible to reveal only the total number of patients affected by a specific disease, without the need to decrypt personal data. Thus, functional encryption can be very interesting in big data cloud storage and has even been presented as the solution to big data security[28]. However practical implementation is still in its infancy due to its slow performance [115].

The above-mentioned techniques are overall aimed at allowing more flexibility regarding access and retrieval of encrypted data. In the next paragraph we review some techniques that go a step further in offering encrypted search and computations.

### 4.2.2 Encrypted search

Search is one the most important computer operations and a core area of databases and information retrieval. Encrypted search can thus be a very powerful tool, especially for big data analytics, allowing full search functionality without the need to disclose any personal data. For example it could be very useful in the context of query – answer systems, where the necessary information is retrieved without accessing the original data.

Although in principle the concepts of search and encryption might seem contradictory, there are today some methods, mostly at research level, that could achieve what we call "searchable encryption". Broadly speaking, the state-of-the-art provides solutions that can achieve different trade-offs between privacy, efficiency (performance) and query expressiveness [116] [117]. Some of these techniques are presented below.

**Property Preserving Encryption**

When maximizing efficiency and query expressiveness, the best available technique today is probably Property Preserving Encryption (PPE) [116][118]. PPE is based on the idea of encrypting data in a way that some property is preserved, e.g. if a>b, enc(a) > enc(b). The simplest form is deterministic encryption (DTE) that preserves equity (if a=b, enc(a)=enc(b)). More complex forms include order-preserving encryption (OPE) and orthogonality preserving encryption. PPE offers fast search features and has been adopted in certain solutions, such as CryptDB [119] and Cipherbase [120]. Still, given the fact that human properties are usually of low entropy (e.g. in medical databases), PPE is vulnerable to data inference attacks and can raise relevant security and privacy concerns [118][121].

**Structured encryption**

When maximizing privacy and efficiency, Boolean keyword search on encrypted data can offer a good option. These types of solutions are based on structured encryption, using symmetric key cryptography or public key cryptography. Still, as mentioned in [117], the limitation of these approaches is the lack of query expressiveness which makes them more suitable for searches in unstructured data, such as web mail and services like Dropbox and OneDrive.

---

[28] See for example: http://blogs.scientificamerican.com/observations/how-to-reconcile-big-data-and-privacy/

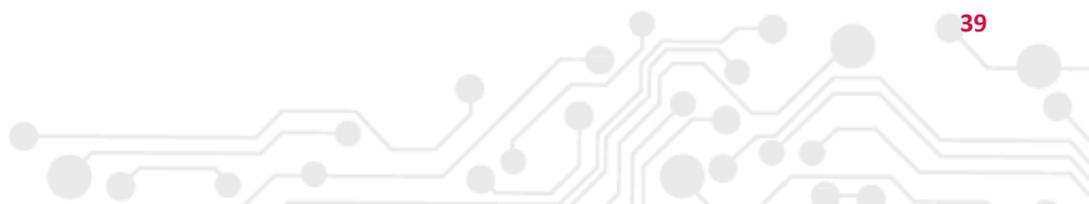

Symmetric Searchable Encryption (SSE) [122] encrypts the database using a symmetric encryption algorithm and allows for later research matching a given keyword. It is appropriate when the entity that searches over the data is also the one that generates it. In particular, a data structure is built that supports fast search on the data set (e.g. an index) and is encrypted using the SSE scheme. The data set is also encrypted using a standard symmetric encryption scheme (e.g. AES). In this way, in order to query the data, it is only necessary to query the encrypted data structure. Specific research in the context of big data has shown the possibility of encrypted search for very large data sets (at terabyte scale) [123] and on distributed databases [124]. In [125] it was shown the possibility of using SSE in correlation rule analysis in medium sized medical databases to predict side effects of certain treatments (and their dependability). In [116] possible advances of SSE for big data are discussed, in particular relating to the design of graph encryption schemes and support of relevant queries.

Public Key Searchable Encryption (PEKS) encrypts the database using public key encryption and allows keyword search[29]. It is appropriate in any setting where the party searching over the data is different from the one that generates it. In contrast with SSE, in PEKS anyone can search a certain keyword on encrypted data and complex searches are possible (so advanced query expressiveness). Still, its efficiency and security are lower than SSE in big databases.

### 4.2.2.1 Privacy preserving computations

When maximizing privacy and query expressiveness the best solutions are Fully Homomorphic Encryption (FHE) and Oblivious RAM (ORAM). These types of techniques fall under the overall area of privacy preserving computations and they are very interesting and emerging research fields. Still, their performance is very low to allow for their practical implementation today, especially regarding FHE. ORAM-based solutions have better performance but are still inefficient in the case of big data [116]. Secure Multiparty Computation (SMC) is another approach, less secure than FHE but with better efficiency in certain types of implementations.

**Homomorphic Encryption**

The word homomorphic has its roots in the Greek language and means "same shape" or "same form". In cryptography its main concept is that analysis can be performed in the ciphertext in the same way as in the plaintext without sharing the secret key (and thus decrypting the data). Beyond the history of homomorphism in cryptography, broadly speaking there are two types of homomorphic encryption: Fully Homomorphic Encryption (FHE) and Somewhat Homomorphic Encryption (SHE). FHE enables an unlimited number of computations but with great efficiency loss, whereas SHE allows a limited number of operations but has better performance than FHE.

One of the issues hindering efficiency in FHE is "noise" that occurs every time an operation on the ciphertext is performed. In 2009 Craig Gentry proposed an approach for FHE, using a process called bootstrapping, aiming to overcome the noise problem and thus achieving a practical implementation of a fully homomorphic system [126]. Still this solution received critics[30] in the sense that the technology is too complex, slow and impractical for cloud computing and big data. For example it has been argued that if the

---

[29] For more details and references, see [6].
[30] See for example Bruce Schneier blog on FSE:
https://www.schneier.com/blog/archives/2009/07/homomorphic_enc.html

Gentry's proposal was used by Google to search the web homomorphically, the normal computing time would be multiplied by about a trillion [127].

It is of course widely acknowledged that a practical FHE solution would greatly contribute to solving security and privacy problems in big data and facilitate adoption of cloud computing services. This is why FHE is a very interesting research field not only for academics but also for the industry [128][129]. Opinions on the possible adoption time of FHE vary: some have predicted that a feasible FHE solution will be available in the next decade, whereas others that it would take at least 40 years before a fully homomorphic system is available [127]. At the current point in time the schemes currently in use in applications are not fully homomorphic: they are either SHE or homomorphic over a limited number of operations (and thus limited to only a few applications) [130][131]. Moreover, most schemes proposed so far are based on public key cryptography, which is not applicable to all use cases in a big data scenario (e.g. a user storing his/her data on the cloud and requiring only a secret key). Also, most current implementations cannot be applied in cases where more than one parties are involved (multi party computations) [132].

**Oblivious RAM**

Another technology that is particularly interesting in this context is that of Oblivious RAM (ORAM) [133]. Its rationale is based on the fact that encryption alone cannot protect data inference, as the sequence of storage locations accessed by the client (the access pattern) can also leak personal data[31]. Oblivious RAM algorithms allow a client to store large amounts of data (e.g. in the cloud) while hiding the identities of the entities being accessed. There is interesting research in the field in order to find practical applicability of the ORAM scheme in large data stores [134][135]. Moreover, beyond its utility in protecting the access pattern in a database, ORAM allows many parties to run secure computations over the data. The main scope of research in this field, is increased efficiency ORAM optimization schemes [136].

**Secure multi-party computation**

Secure multi-party computation (also known as secure computation or multi-party computation - MPC) is a field of cryptography aimed at enabling different parties to compute a function over their inputs, without disclosing the individual inputs. For example, if three persons x, y, z want to find out who has the highest salary (by computing max(x, y, z)) without revealing to each other their individual salaries. This very basic scenario can be generalised to where the parties have several inputs and outputs, and the function outputs different values to different parties [137].

There are many approaches to SMC, including different cryptographic tools, such as zero-knowledge proof, oblivious transfer, Yao's millionaire protocol, etc [6]. In the context of big data and cloud environment, MPC offers weaker security than FHE, especially when many untrusted parties are involved. In such a case, although each party cannot learn anything from the data, it has been shown that if sufficiently many parties are corrupted by an adversary and pool their information, they can break confidentiality [138]. Other open issues of MPC in big data are listed in [139]. Still, MPC can be quite efficient in certain practical implementations, as was shown in the cases of Danish farmer sugar beets [140] and the smart metering deployments in the Netherlands [141].

---

[31] For example, it has been shown that by observing accesses to an encrypted email repository, an adversary can infer as much as 80% of the search queries, [230].

## 4.3 Security and accountability controls

Security is a fundamental part of privacy protection and, thus, central for big data. In order to achieve an appropriate level of security, several measures need to be in place at different levels of the big data value chain. Still, "traditional" approaches to information security fail in big data, as they are usually intended to address small-scale static data. According to Cloud Security Alliance, the high-priority security and privacy issues that need to be further researched in big data include [142]:

- Secure computations in distributed programming frameworks
- Security best practices for non-relational data stores
- Secure data storage and transaction logs
- End-point input validation/filtering
- Real-time security/compliance monitoring
- Scalable and composable privacy-preserving data mining and analytics
- Cryptographically enforced access control and secure communication
- Granular access control
- Granular audits
- Data provenance

Each of the above-mentioned issues can be a research topic in itself and they are all (alone and in combination) essential for the security of personal data. In the context of this report we are focusing on privacy preserving analytics and, in this respect, we have already discussed to a certain extent relevant aspects, in particular anonymization and encryption. Following the CSA priority list, we also address in the next paragraphs some specificities of the key security big data issues in relation to privacy and data protection. More information on security good practices and recommendations in big data systems can be found in ENISA's reports on big data threat landscape and big data security[32].

### 4.3.1 Granular access control

Access control is one of the most fundamental security measures that is applicable to any application, ensuring that only authorised processes and procedures can gain access to data. In the case of big data, where data are characterized by large diversity (ranging from structured to semi-structured and unstructured) and different level of privacy requirements, traditional approaches to access control, like role-based access control or access control lists are becoming unmanageable.

To this end, there are approaches that can conceptually support fine grained access control policies in big data based on attributes that are evaluated at run-time, such as Attribute Based Access Control (ABAC) [143]. Rather than just using the role of a user or (more generic) of a data set to decide whether or not to grant access, ABAC can make a context-aware decision through the combination of several attributes. The rules based on these attributes can also embody contextual privacy requirements as outsourcing restrictions, data minimization and purpose specification. eXtensible Access Control Markup Language (XACML) [144] is a well acknowledged open standard that supports the creation of access control policies and comparative evaluation of access requests according to predefined policy rules.

Still, it is important to note that fine-grained access control has a large administration overhead which is usually too high for such schemes to be practical today. Moreover, the use of attributes, despite the original privacy preventive concept has some further implications and could potentially lead to profiling, depending on the overall context in which they are being used (and/or reused in the case of big data) [145]. Further research is needed both regarding the practical application of ABAC in big data, as well as its overall social and data protection impact.

---

[32] See ENISA's web site for latest publications, https://www.enisa.europa.eu/

### 4.3.2 Privacy policy enforcement

As also described in CSA's report [142], the auto-tiering process in big data cloud solutions allows for automated moving of data between different tiers, which provides efficiency and cost management. Still, this would result in scenarios where critical information (e.g. health data stored for research) would be transferred to lower level security tiers (if for example this data are not often used). This points to a more general privacy problem in the context of big data: moving, sharing and transferring of data between different systems (and different data controllers) could result in lowering the protection level of personal data, even if this level is high at the initial point of collection and processing. Therefore, security and privacy policies, even if in principle available, might be lost and/or neglected in the course of the big data value chain.

Automated security policy enforcement mechanisms can be central in this respect and there are already relevant features in existing big data solutions, especially related to data expiration policies[33]. Automated scanning of data and logs (and subsequent deletion based on security policies) is offered in most database systems today. There is also interesting work in the field, based on virtualisation and trusted computing[34]. In particular, trusted computing uses tamper-resistant hardware storage and machine-readable policies and encrypts data twice: the outer level can be decrypted only by trusted hardware, whereas the inner level can be decrypted by software that has been checked and meets the policy requirements[35]. One disadvantage of this approach is that all entities need to be part of the system (and, thus, trusted), which is not always feasible in a big data scenario.

Enforcing specific privacy settings as part of these mechanisms is an additional challenge, for example regarding access control policies, purpose limitation, data subject's consent and location of data. The use of semantics, policy languages and metadata annotations are useful tools in this respect and they have been applied also in big data cloud environments [146][147][148] (see also section 4.5. on sticky policies). Still, interoperability is a critical issue in this respect, in order to allow for the technical implementation (enforcement) of such policies. Also, it is important to combine these mechanisms with proper information to the users regarding the processing of their personal data.

### 4.3.3 Accountability and audit mechanisms

Accountability is a key concept in data protection and is going to be reinforced in the context of the proposed General Data Protection Regulation. From a technical point of view, it is very much related to the implementation and enforcement of privacy policies, in particular ensuring that such a policy is in place and being able to demonstrate that it is actually respected. In order to provide for an accountable system it is, thus, important to have automated and scalable control and auditing processes that can evaluate the level of privacy policy enforcement against the predefined machine-readable rules, as well as to provide evidence for any misconduct. Different types of measures can contribute to this, e.g. logging and monitoring controls, which form part of most big data solutions today.

---

[33] See for example MongoDB's "time to live" or TTL collection feature, https://docs.mongodb.org/manual/tutorial/expire-data/
[34] See for example [231] on automated security policy enforcement in multi-tenant virtual data centres.
[35] http://www.trustedcomputinggroup.org

Still, distinct measures alone are not enough and the need for a thorough automated accountability and audit framework for privacy is an emerging and very interesting field [149], based on a variety of technologies such as formal accountability models, computer systems design and cryptographic techniques[36]. According to [150], an accountable system should include automated policy compliance tools, provenance management, detection of violations, complete audit logs, policy awareness and redress mechanisms. It is interesting in this respect to mention the A4Cloud project that claims to deliver an integrated accountability framework for security and trust in cloud services by the development of mappings between contracts/SLAs and evidence gathered from logging and user-centric tools[37]. Of course, automated accountability mechanisms are still far from full implementation. Related open issues include the level of abstraction of the policies, automated policy conflict resolution, usability aspects, etc.

Automated tools should be combined with other accountability mechanisms, such as Privacy Impact Assessments and personal data breach notifications, which need to be further analysed and developed for big data cloud environments.

### 4.3.4 Data provenance

There are various definitions of data provenance, connecting it with the ownership, custody and location of information [151]. In the context of big data, where analytics transform raw and distributed data into useful and meaningful outputs, provenance can attest data origin and authenticity, qualify assertions and justify unexpected results. It also forms integral part of an audit-compliance-accountability process, as described above. This can be useful both for the data analyzer, but also for the data subject, as provenance mechanisms could allow him/her to track how the data are being processed [152].

However, the inclusion and transmission of provenance data (or metadata), even after several transformations can result to the unwanted disclosure of personal data. For example, an individual tracing his/her own data may accidentally access data of other persons if they are part of the same process. As another example, metadata related to health information (e.g. healthcare institution, time and date) might be enough to single out certain persons, in combination with other information (e.g. camera logs). The issue in data provenance is that, unlike statistical analysis, no aggregation takes place and the data shown are exact. Access control and query/answer systems can be a possible solution [153], although it is a challenge to find the right combination between utility and privacy.

According to CSA the threats to provenance information can be faced by securing provenance collection, pairing a fast, lightweight authentication technique, and fine-grained access control of provenance [154]. Recently the notion of provenance-based access control (PBAC) has been formulated and a base PBAC model (PBAC) together with an underlying provenance data model has been specified [155]. It has been shown that, unlike Role-based Access Control, PBAC directly maintains and utilizes the necessary information for dynamic segmentation of duties enforcement [156].

## 4.4 Transparency and access

Proper information and transparency is a key issue in any data processing, so as to allow individuals to understand how their data are being processed and to make relevant informed choices. In big data transparency is needed more than ever, taking into account that analytics are in many cases not based on information that individuals knowingly provide about themselves, but on data observed or inferred from online activities, locations, smart devices, etc. [4]. Therefore, transparency needs to expand beyond the original point of data collection and individuals should be adequately informed about the logic and the criteria applied in the context of big data analytics automated decision-making processes.

---

[36] See http://dig.csail.mit.edu/2014/AccountableSystems2014/ for a comprehensive approach towards accountable systems and relevant research on the field.
[37] http://www.a4cloud.eu/

To this end, new information and transparency models need to be developed. Purely textual information does not seem to cope with the evolution of services and to comprehensively inform users on the processing of data occurring in the complex big data value chain. Many studies have pointed out the ineffectiveness and time consumption of this approach to inform users. For example in [157] it was shown that, in order for an average user to read the privacy policies for all visited web services, he/she would need to spend approximately 30 working days per year.

To improve the effectiveness of information, multichannel and layered approaches have been suggested [158], which can provide information to the users at different stages of the processing and at incremental levels of detail. Usability is a key aspect in this approach, in the sense that layered information should also be presented in plain language and simple notices (in contrast with the long privacy policies of most web sites today). Technologies that can provide visual representations and graphic information (like immersive computing [159]) can be also very interesting in this respect.

Standardized icons and pictograms is another promising emerging approach for transparency in big data. The main idea is to use icons in order to offer an immediate understanding of the privacy policy of a given web site, without the need to read any text. Disconnect Privacy Icons[38] add-on is an example of such a tool, where specific icons can inform the user of the expected collection and use of his/her data when visiting a web site, the data retention, if do-not-track is honored, if the user's location is tracked, if the website is vulnerable to the Heartbleed attack, etc. In Disconnect's approach the icons are generated by analyzing the site's privacy policy and, as has been presented, many widely used service providers have received low rankings [160] Another interesting initiative is that of Mozilla privacy icons [161], proposing a set of icons with information about possible behavioral advertising, reuse of data, use by law enforcement, etc. Still, one downside of the icons' approach is that a web site would probably not be willing to adopt negative icons and, thus, reveal possible privacy intrusive policies. For example if the data are shared with other parties or used for advertising, the controller would probably refrain from displaying a relevant icon. It has been mentioned that a solution to this problem could be to make the icons machine readable, in a way that if an icon is not found, then the negative option would be adopted. Still, the use of icons cannot be enforced and trust is an open issue[39].

Along with transparency (or as part of it), providing access to users on their data is an important privacy condition as well as an obligation of data controllers. However, users rarely exercise their right of access to online providers, either because they are not aware of it (lack of proper information) or because it is very difficult (no online button or other type of automated process). Data portability, closely linked to data access, is an additional and important tool for users, giving the possibility to change service providers without losing their data. There are already some interesting initiatives in this respect [162], for example the USA Smart Disclosure project of the National Science and Technology Council, as well as the Green button initiative allowing easy access to energy consumption. Another relevant project is the Midata UK initiative [163], providing access in transactions and consumption for the energy, finance, telecommunications and retail sectors. Also, the French MesInfos platform for access to financial, communication, health, insurance and energy data [164]. Still, easy and practical data subjects' rights management is far from being reality today and there is need for more innovation and creativity in this domain, especially in the area of semantics and authentication, taking into account that transparency and access can empower users, also for the benefit of big data analytics.

---

[38] https://disconnect.me/icons
[39] See for example a relevant discussion in: http://www.webmonkey.com/2010/12/new-privacy-icons-aim-to-save-you-from-yourself/

## 4.5 Consent, ownership and control

User control is a crucial goal in big data and it can be reached through a multichannel approach. Consent is one possible solution (and probably the most prominent one). Other methods and tools can also contribute by ensuring accurate audits and determining the compliance of controllers and processors with the rules. An example of such a method is "tagging" every unit of personal data with "metadata" describing data protection requirements. This is also the perspective of semantic web, but putting tags and rules on data is a costly activity which will require a multi-stakeholder effort. Tools that put the data subject in charge of managing their data is also a promising and emerging research field.

### 4.5.1 Consent mechanisms

The continuous repurposing and making use of already processed or inherent data sets, has made the traditional consent models insufficient and obsolete in big data. This has led to many arguments against the very concept of consent. Still, consent is a fundamental data protection element and, like many other, it has to adapt to the new technological landscape with new usable and practical techniques.

In fact, although consent is often perceived as obstacle to service usability, consent mechanisms have been very much engineered by the industry in the recent years and collecting consent does not constitute always a real barrier for the usability of a service, as it has been shown by the recent Google consent policy[40]. Nowadays, its unacceptance seems to be more a psychological barrier. User friendly consent mechanisms have also been proposed by Data Protection Authorities, mainly based on engineered banner solutions[41].

Practical implementation of consent in big data should go beyond the existing models and provide more automation, both in the collection and withdrawal of consent. Software agents providing consent on user's behalf based on the properties of certain applications could be a topic to explore. Moreover, taking into account the sensors and smart devices in big data, other types of usable and practical user positive actions, which could constitute consent (e.g. gesture, spatial patterns, behavioral patterns, motions), need to be analyzed.

### 4.5.2 Privacy preferences and sticky policies

Another approach that can be quite interesting in providing user control is offering the possibility to both controllers and data subjects to express their privacy policies and requirements prior to the processing, an approach also known as "privacy preferences". Sticky policies can provide a mechanism for attaching privacy preferences to specific data sets and accordingly drive data processing decisions [165]. In particular, through the formal expression of acceptable privacy preferences, data subjects are able to communicate to each data controller their inclinations and dispositions regarding the processing of their data. For example they can indicate the acceptable purposes on the processing of their data, the allowed recipients, deletion periods, etc. Following the formalization of the data controllers' commitments, through privacy policy documents, on the purposes and conditions of the processing, formal statements from both parties can be compared prior to the collection of personal data. The outcomes of this comparison can be regarded as a privacy contract between the two parties, detailing all the agreed instances of data collection and further processing.

---

[40] http://www.google.com/about/company/user-consent-policy.html
[41] See for example the relevant proposals of the Italian DPA in
http://www.garanteprivacy.it/web/guest/home/docweb/-/docweb-display/docweb/3167654 and also in
http://www.garanteprivacy.it/web/guest/home/docweb/-/docweb-display/docweb/3295641.

Through the exploitation of EXtensible Markup Language (XML), P3P standard (Platform for Privacy Preferences) and P3P Preference Exchange Language (APPEL) [166], this approach can enable users to keep control over the collection, use and sharing of their personal data with various data controllers. In addition to the "take-it-or-leave-it" model, several techniques have been proposed that also support automated negotiation procedures between the data subject and the data controller [167][168][169]. Data subjects define a set of privacy preferences which sets out how negotiation will be performed. Through different negotiation steps, a commonly acceptable solution for conflicting privacy needs is reached through a fine-grained privacy contract that governs the use of personal data during the specific instance.

The enforcement of privacy preferences and sticky policies is very important for their practical adoption and implementation. In principle enforcement is weak, as the individual would need to trust that all parties involved in the processing of personal data will respect his/her preferences. The use of encryption techniques to enhance sticky policies enforcement is a promising research area to this end: if the data subject can define his/her privacy preferences in the encryption key, then the right recipients can be targeted without the need to trust every involved party. Different encryption methods can be applied. Public key encryption (for shared key exchange) combined with symmetric key encryption is one possible solution. Identity-based encryption can enable the data subject to express his/her policies by choosing any string as the public key. Research on attribute based encryption (key-policy ABE and ciphertext-policy ABE) also exists, as well as on proxy-re-encryption (especially on access control policies) [170]. The role of one or more trusted authorities is central in all different approaches for key distribution and management.

Despite the possibilities that privacy preferences and sticky policies can offer in the context of big data, it should be mentioned that the proposed model cannot be a substitute of consent, but rather a facilitator for user's overall control and choice. Indeed many of the critics especially on P3P are regarding the poor consent implementation and the fact that policies can be quite coarse, leading to excessive data disclosure. Alterative research solutions, such as S4P [171] [6] , need to be further explored and practically implemented. Moreover, further research is needed in the context of big data, especially regarding mediation of conflict resolution of heterogeneous policies, auditing and compliance for policy enforcement.

### 4.5.3 Personal data stores

A range of technical solutions have been proposed to give data subjects' increased control over their data through transition from distributed data models to user-centric models. As acknowledged in a study on personal data stores conducted on behalf of European Commission by Cambridge University [172], all such technologies whether called personal data vaults, personal data lockers, personal clouds, or, personal data stores (PDS), "enable individuals to gather, store, update, correct, analyse, and/or share personal data." Based on this concept, data subjects have full control over the release of their personal data and retain ownership even after the release, contrary to existing systems where they have limited-to-no access or control.

The development of personal data stores can be envisaged with the use of the semantic data interchange protocol [173] and relevant trust frameworks that allow for embedding the privacy preferences of the users into link contracts between the user and the service provider [174].

The ability to grant and withdraw consent for access to personal data could transform the consent procedure during big data analytics: instead of relying on existing data sets, data controllers could request access to specific personal data directly from the data subjects. This empowerment, though, comes along with obligations and the additional burdens of reviewing numerous requests and being liable for managing access credentials. Due to the centralisation of data (either physically or through a common interface), security risks also evolve, as the PDS could be regarded as a single point of failure.

An example of PDS research projects is Personal Data Vault [175], which has explored three different mechanisms to automatically manage data policies and has been evaluated using real world stream data. Databox [176] is another tool in this direction, as well Virtual Individual Servers (VISs), which has been evaluated against centralized third-party services and in sharing data directly from mobile devices [177]. OpenPDS [178] is a personal data store focusing on big metadata from mobile phone, credit card, browsing history. It "allows individuals to collect, store, and give fine-grained access to their metadata to third parties" and proposes auditing and security mechanisms such as attacks by well-behaved and malicious apps or attempts at compromising the host. It is based on a mechanism called SafeAnswers and lets third-parties ask questions and get answers from individual data without seeing or directly accessing the raw metadata. Enigma [179] combines openPDS's SafeAnswers mechanism with an optimized version of secure-multiparty computation and a blockchain to manage access control, identities, and event logging.

Examples of commercial solutions that do exist today are Higgins[42], Mydex[43] (with focus on text, numbers, images, video, certificates and sound), and You.tc[44] mainly on social network posts and credit card transactions. CyberAll is a relevant project developed by Microsoft early in 2001 to centralize and encode people's data [180]. Persona focuses on giving user more control over their social network data, allowing users to hide their data using attribute-based encryption (ABE) [181]. Confab [182] is a toolkit to design privacy-sensitive ubiquitous computing applications allowing users to selectively share information.
While these projects differ in the type of data they handle (files vs data), how they are implemented (centralized, decentralized), how much control the user has, as well as in their dissemination model (personal or rented virtual machine, as a service, through current data holders such as mobile phone carriers or trusted third-parties), they all focus on giving individuals control over their (big) data, including more transparency on how data are used.

---

[42] http://www.eclipse.org/higgins/
[43] http://mydex.org/
[44] http://you.tc

# 5. Conclusions and recommendations

This report aims at adopting the concept of privacy by design in the era of big data by examining the potential and applicability of specific privacy enhancing technologies in the different phases of the big data value chain. Our overall aim is to support the goal of "big data with privacy", contrary to the often presented battle of "big data versus privacy", putting technology at the core of the discussion and at the service of data protection.

One straightforward conclusion in that respect is that the road to this goal is still long and hard. Big data are becoming bigger every day, whereas privacy preserving mechanisms are not by design or by default in the heart of analytics. Unfortunately it is only after wide scale data breaches that service providers and users start understanding and re-evaluating the importance of embedding privacy protection as a core element of big data uses and applications. And even this takes time: it is up to the moment that big data becomes "personal" that users start questioning the free online services and the sensors-based automation and that trust to the new data-driven environment breaks. Still, this study argues that the goal of "big data with privacy", although difficult, is not unattainable. On the contrary, it suggests that big data's prosperity needs the privacy principles to regain users' trust and provide for better analytics, improving our everyday life, without invading our personal sphere.

To this end, with this report, we aim at taking a first look in the way that technological advancements for big data can meet and integrate technological advancements in privacy. This should not be considered as an exhaustive presentation of all available and possible techniques, but rather as an attempt to take this discussion a step forward and to engage all relevant stakeholders in a more humanistic data-protection driven analytics development.

**Privacy by design applied**

Moving ahead from the generic concept of privacy by design, the big data industry and the data controllers need to embed specific privacy and data protection measures into their processes and systems. But what does this mean in practice? As already mentioned, the proposed General Data Protection Regulation has a new obligation for data controllers and processors for data protection by design and by default. How can this obligation be embodied in the complex big data landscape, where different controllers continuously interact in a way that even one can reduce the privacy level for all?

In this report we tried to explore the privacy by design strategies in the different phases of the big data value chain. However, every case is different and further guidance is required especially when many players are involved and privacy can be compromised in various points of the chain. Privacy Impact Assessments (PIA) can be valuable tools to this end but, again, more work is needed to address the various and complicated analytics environments, taking into account all the relevant stakeholders. Specific examples and solutions, including the results of PIAs, need to be openly presented and discussed, so as to provide useful use cases for similar data processing scenarios.

*Data Protection Authorities, data controllers and the big data analytics industry need to actively interact in order to define how privacy by design can be practically implemented (and demonstrated) in the area of big data analytics, including relevant support processes and tools.*

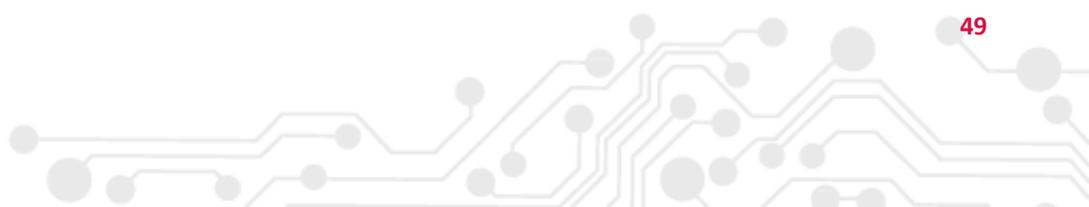

**Decentralised versus centralised data analytics**

Big data today is about obtaining and maximizing information. This is its power and at the same time its problem. Following the current technological developments, we take the view that selectiveness (for effectiveness) should be the new era of analytics. This translates to securely accessing only the information that is actually needed for a particular analysis (instead of collecting all possible data to feed the analysis). Having said that, we recognise that this might not be possible to all cases and the "traditional" analytics scheme would continue to exist. Still, such a shift from centralization to de-centralized analytics environments could be beneficial in many cases, for example when the type of the analysis is known and the data are coming from many distributed sources. As has been shown in this report, there are already techniques, both in the fields of anonymization and cryptography, which can offer interesting solutions to this end. Moreover the increased computing processing power today can support decentralised models through fast communication channels. Research is still needed to provide for a practical and viable implementation of such techniques into real case scenarios.

*The research community and the big data analytics industry need to continue and combine their efforts towards decentralised privacy preserving analytics models. The policy makers need to encourage and promote such efforts, both at research and at implementation levels.*

**Support and automation of policy enforcement**

Developing a proper privacy policy with respect to underlying legal obligations is central in any data processing scenario. In big data an additional challenge is introduced: in the chain of co-controllership and information sharing, certain privacy requirements of one controller might not be respected by another. In the same way, privacy preferences of the data subjects (which may form part of the data controller's policy in the first place) may also be neglected or not adequately considered. Monitoring of compliance is not always feasible, for example in cases where audits from the EU Data Protection Authorities are limited due to the location of the data or the complexity of the systems used. Therefore, we find that there is need for further research and use cases in the area of automated policy definition and enforcement, in a way that one party cannot refuse to honour the policy of another party in the chain of big data analytics. Semantics and relevant standards, as well as cryptographically enforced rules, are fields that require careful study, taking also into account the legal framework for the protection of personal data.

*The research community and the big data analytics industry need to explore the area of policy definition and to embody relevant mechanisms for automated enforcement of privacy requirements and preferences. The policy makers need to facilitate the dialogue between research, industry and Data Protection Authorities for effective policy development and automation models.*

**Transparency and control**

Empowering end users is an essential way to get them actively involved for their own benefit, but also for the benefit of big data analytics. In the age of big data, "traditional" notice and consent mechanisms fail to provide proper transparency and control. In the context of this report, we discussed some interesting concepts and ideas, such as the privacy icons, sticky policies and personal data stores. Still, the very idea of consent needs to be reinforced (instead of being opposed or presented as a burden). New technologies, require new consent models and the industry needs to be open and creative in providing practical opt-in solutions, complemented by opt-out and access mechanisms that can put users in control throughout the data processing chain.

*The big data analytics industry and the data controllers need to work on new transparency and control measures, putting the individuals in charge of the processing of their data. Data Protection Authorities need*

*to support these efforts, encouraging the implementation of practical use cases and effective examples of transparency and control mechanisms that are compatible with legal obligations.*

**User awareness and promotion of PETs**

End users also need to take action, not only by demanding transparency and control, but also by responsibly protecting themselves. There are already numerous privacy enhancing tools for online and mobile protection, such as anti-tracking, encryption, secure file sharing and secure communication tools, which could offer valuable support in avoiding unwanted processing of personal data. Still, more work is needed in evaluating the reliability of these tools and their further applicability for the general public. ENISA is actively working in this field, in collaboration with other interested stakeholders. User awareness is key in this respect, in order to adequately and actively involve all ages and user groups in using responsibly and proactively the various online and mobile services that feed the big data analytics.

*The research community needs to adequately address aspects related to the reliability and usability of online PETs. The role of the Data Protection Authorities is central in user awareness and promotion of privacy preserving processes and tools in online and mobile applications.*

**A coherent approach towards privacy and big data**

This report focused mainly on technology for big data privacy. Still technology alone is not enough. The proposed General Data Protection Regulation will provide for new rights and obligations that need to be practically implemented, also in relation to the overall European Commission's vision and research programme for big data. Opposing or vague policy requirements can only make the picture blurry, undermining both individuals' privacy and big data quality of results. To this end, big data research should include as a core element the integration of privacy enhancing technologies. And research on privacy enhancing technologies should take into account the dimension and emerging big data landscape.

*Policy makers need to approach privacy and data protection principles (and technologies) as a core aspect of big data projects and relevant decision-making processes.*

ENISA will continue working in this direction, with main focus on addressing the challenges of (big data) technology with the opportunities of (privacy) technology for the benefit of all involved stakeholders.

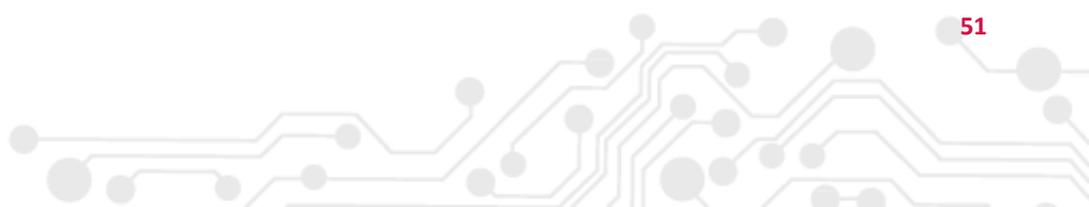

# Annex 1 - Privacy and big data in smart cities: an example

Big data analytics is already happening today and in many sectors of our everyday life. In order to demonstrate the big data analytics (and the relevant privacy considerations) in practice, in this section we focus particularly on the area of smart cities, defining and further analysing three different analytics scenarios and their aggregation in the big data pool. The selection of smart cities as an example was twofold: a) first, it is a fast developing area of analytics that involves several cases of personal data processing, and b) second, it concerns and affects a great (and increasing) number of diverse population in Europe.

To this end, for the scope of this report, we present the use cases of smart-parking, smart-metering (smart grid) and citizen platform, together with their interconnection in the context of a smart city. Following a short overview of the application scenarios, the focus is shifted on the processing of personal data in analytics, the induced privacy risks and the relevant risk mitigation strategies.

It should be noted that it is not our purpose to perform a legal analysis of each particular use case regarding the processing of personal data. On the contrary, our study only aims at providing a first insight on the privacy risks arising from each specific application and (mainly) from the correlation of data from different applications. Moreover, following the analysis of Chapters 2, 3 and 4 we aim at showing in practice how some of these techniques can contribute to the protection of personal data in big data analytics.

## A.1   Introduction to smart cities

According to the European Commission [183] "A smart city is a place where the traditional networks and services are made more efficient with the use of digital and telecommunication technologies, for the benefit of its inhabitants and businesses. The smart city concept goes beyond the use of ICT for better resource use and less emissions. It means smarter urban transport networks, upgraded water supply and waste disposal facilities, and more efficient ways to light and heat buildings. And it also encompasses a more interactive and responsive city administration, safer public spaces and meeting the needs of an ageing population".

It is obvious that ICT technology plays a vital role in the realization of the smart cities vision. According to the ITU-T Focus Group on Smart Sustainable Cities [184], ICT acts as the platform to aggregate information and data to help enable an improved understanding on how the city is functioning in terms of resource consumption, services, and lifestyles. To achieve this it is obvious that large amounts of structured and unstructured data are collected and further processed from a diverse set of applications, throughout all city activities, including data from sensors, mobile networks, social platforms, etc. Big data analytics is core in using all these data to improve the functioning of the city, provide new services and tools and overall increase the quality of life of the citizens. Of course, at the core of this data, are the individuals' (end users/citizens) personal data collected through various devices such as smart meters, smart phones, connected vehicles, etc.

During the last years, the European Commission launched specific research activities, under the 7[th] Framework Programme, related to Smart Cities and the potentials of ICT[45]. The expected outcomes of these research projects were to facilitate wider uptake of IoT-based systems with an emphasis on sustainable smart city applications, while promoting technological focus on built-in privacy, security and on scalable data management.  Example of such projects include ALMANAC (Reliable Smart Secure Internet of Things for

---

[45] Under the ICT-2013.1.4 objective: A reliable, smart and secure Internet of Things for Smart Cities, http://cordis.europa.eu/fp7/ict/docs/ict-wp2013-10-7-2013-with-cover-issn.pdf

Smart Cities)[46], SMARTIE (Secure and Smarter cities data management)[47], RERUM (Reliable, Resilient and secure IoT for smart city applications)[48], and CityPulse (Real-Time IoT Stream Processing and Large-scale Data Analytics for Smart City Applications).[49] Even through different approaches, proposed frameworks and technologies, all of these projects aim at providing interoperability amongst heterogeneous architectures, automation of data flows, and employment of data analytics to support near real time processing of large volumes of diverse data. ENISA has also been working on cybersecurity and smart cities, especially in the area of public transport and smart grids.

As already mentioned, for the purpose of this report, we focus on three specific use cases in the context of smart cities, i.e. smart parking, smart metering, and citizens platform, and their privacy related considerations.

## A.2 Smart city use cases

### A.2.1 Smart parking

Looking for an available parking space is a daily routine for drivers in big cities, which not only takes up a lot of their time but also has an impact on multiple aspects of the cities. In particular, the increased emission of gases deteriorates atmosphere's pollution and grows traffic congestion, further impacting the effectiveness of surface transportation and fuel consumption. To address this problem, city administrations are deploying IoT solutions which, based on low cost sensors, can infer if a parking positon is occupied (typical sensors are magnetic, passive IR, ultrasound, and cameras).

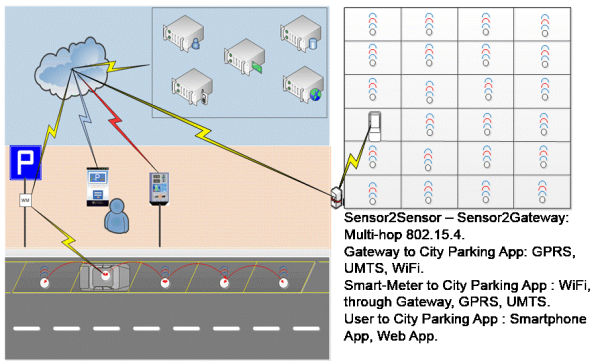

**Figure 1: Smart parking application scenario**

Through real-time data collection and communication, the central system processes the sensors' measurements and, using advanced analytics, it combines occupancy information with the drivers' location sent via a smartphone application. In this way, the drivers are directed towards the parking spot that best matches their needs. The smartphone application might have additional features for the drivers to declare specific preferences, as well as to perform electronic payments, e.g. for parking spaces or other services offered on the spot.

As depicted in Figure 1, in such a scenario many stakeholders may be involved, for example the city administration, the service provider, the telecommunication provider, the bank, an aggregator/mediator and the user (driver). One stakeholder might have more than one roles, e.g. the telecommunication provider might also be the aggregator/mediator, the city administration might also be the service provider, etc.

In order for the smart parking application to be effective, efficient and sustainable, the users (drivers) must disclose either prior or during the provision of a service, a number of personal data. First of all, for the real time identification of nearby parking slots, the location of the user must be sent (via the smartphone) to the service provider. Moreover, the user might wish to declare certain preferences, e.g. maximum acceptable distance of the parking slot, time availability of slot, cost of parking per hour, need for electrical outlet in case of electrical car, etc. If the user wishes to utilise the billing feature of the smartphone application, financial data are also transmitted in the context of the processing (probably via the user's bank). The combination and/or further analysis of these data could also provide additional information, e.g. about the user's everyday habits or contacts (if he/she often parks at a particular place at a certain time of the day).

---

[46] http://www.almanac-project.eu/

[47] http://www.smartie-project.eu/

[48] https://ict-rerum.eu/

[49] http://www.ict-citypulse.eu

The data processing scenario requires the interaction of many data controllers and processors and the scheme varies depending on the assigned roles and responsibilities. For example the city administration is the data controller offering the service, probably via outsourcing to the service provider (processor). The telecommunication operator is also a data controller providing the location data. The bank is a data controller offering the billing system. Each of these controllers processes the data for its own purposes and the aggregation of the data, in the context of the smart parking application, is performed by the mediator/aggregator.

A.2.2 Smart metering (smart grid)

Smart grids are energy networks that can automatically monitor energy flows and adjust to changes in energy supply and demand accordingly [185]. Through the flow monitoring they can facilitate a smoother introduction of renewable sources of energy by accumulating their contribution to the grid, offering more stability to the electrical network and allowing grid operators to balance their networks. When combined with smart metering systems, smart grids can also provide information on real-time consumption.

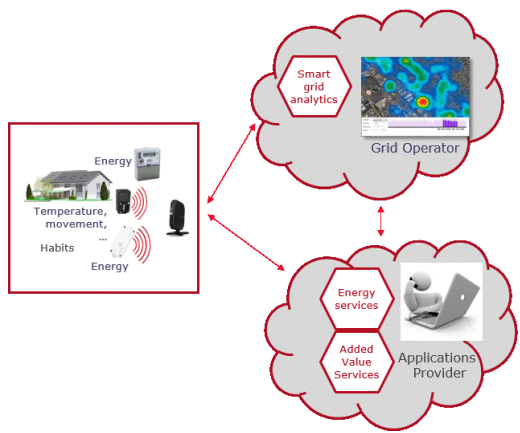

**Figure 2: Smart metering application**

Smart metering systems are smart sensor devices that are connected between the grid edge and the home electrical cabinet and measure the inflow of energy from the grid to the house. This inflow corresponds to the energy consumed within the household. Smart metering systems are connected via a telecommunication network to the metering provider, which processes the data and calculates the total cost of consumption. More advanced meters can be considered as gateways, in which a number of sensors are connected (e.g. smart plugs to monitor individual devices, smart actuators to control devices, temperature and weather sensors to compensate for fluctuations in the environment, inductive sensors to the home cabinets to monitor individual electrical lines within the house, etc.).

The stakeholders involved in such a scenario include the energy supplier, the smart meter/metering service provider, the telecommunication provider, the billing provider and the user. Again one stakeholder may have multiple roles, e.g. the energy supplier may also be offering the smart metering system.

Smart meters are based on the collection, transmission and further use of energy consumption data at a detail level much higher than that of a "traditional" meter. These data also include user's personal data that can be inferred from the energy consumptions patterns. In particular, by analyzing the energy consumption of specific appliances/devices, the user's presence and behavior in the house can be guessed. Moreover, when the smart meter is connected to other sensors, a detailed behavioural profile of the user could also be built (e.g. what time he/she goes to sleep or wakes up, what type of TV movies he/she is watching, etc). This potentially intrusive characteristics of smart metes have raised several privacy concerns [186].

Like in the case of smart parking, several data controllers are involved and interrelated also in the smart metering scenario, such as the energy provider, the service provider, and the telecommunications operator.

## A.2.3 Citizen platform

City administrations are promoting new ways to communicate, engage and interact with their citizens in order to be aware, in real time, of their needs and concerns, and to timely respond and deliver tailored services. One prominent way among smart city infrastructures, to facilitate this interaction, is through mobile crowdsourcing applications (hereinafter citizen platforms).

The scope of these types of applications is threefold: a) empower citizens to report problems that they encounter and request immediate reaction from the city officials (e.g. traffic, waste collection, water outage, emergencies, environmental violations, noise, etc.), b) provide personalized recommendations through city guides, entertainment and shopping suggestions based on citizens' preferences, and c) e-participation services, allowing citizens to give their opinions regarding aspects of the city's life. Citizen platforms usually support interoperability with social media providers, enabling users to tag, comment and share relevant information and content.

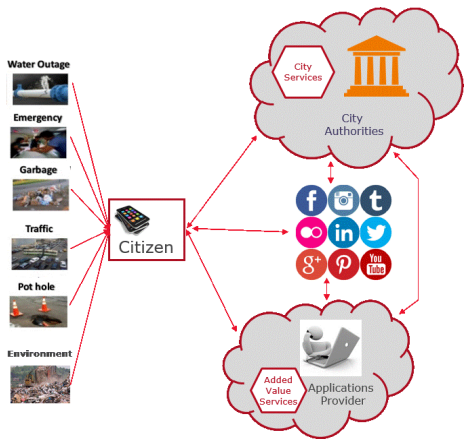

The identified stakeholders in this scenario are the citizens (users of the application), the city authorities, the telecommunication providers, the application providers and the social media providers.

Again, the processing of users' personal data is central in this application, including location information, preferences, habits and beliefs. Due to the connectivity with social media platforms there is a chance that further personal data are inferred in the context of such a processing, for example by combining the user's voting patterns in the citizen app with relevant postings in social media.

**Figure 3: Citizen platform**

There are a number of city applications already deployed today in a number of European cities [187][188] [189][190][191].

## A.3 Privacy considerations

Each of the above-presented scenarios have their own privacy risks that need to be appropriately managed when defining the conditions for the processing of personal data. These risks, however, can grow exponentially if data from the different use cases are correlated in the context of advanced analytics. Such correlation could be very interesting, so as to identify different behavioural patterns and graphs of citizens' movements and activities in the city (conducted for different types of purposes, e.g. demographics, provision of new services, advertising, etc.). However, if not adequately safeguarded, this type of analysis can also lead to very detailed insight in the individuals' life, including everyday movements, habits, social contacts, leisure, etc. This can be a quite typical data analytics scenario, which is not so far from being reality.

To this end, the first issue which immediately emerges in all described use cases is *control*: who is responsible for what. The diversity and complex interrelation of the different data controllers and processors in each of the presented scenarios (and even more in their integration) can create confusion and vagueness. Without a proper data protection responsibility allocation, each actor will be tempted to utilize the data according to its own economic interest, overriding users' preferences. Potential *reuse of data* without the users' information and consent cannot be excluded. This is not just a theoretic scenario, but it is already relevant in the so called ("old technology") value added services, and the most disturbing effect is the undesired activation of new services which generates costs and annoyance for users.

Another risk coming from the dispersion of the roles is the potential *duplication of data*. Again, if responsibilities are not properly defined before the service is delivered to users, each actor will be tempted to store the data on its systems for its own purpose (mainly optimization and engineering), and very likely with different level of security attached to each data reservoir, depending on the economic resources and data protection culture. It is worth reminding that in an interconnected system, the *level of security* is set by the weakest link, with consequences on the probability of occurrence of data breaches.

Also, and still connected to the dispersion of roles, specific risks emerge with the effectiveness of the privacy policies adopted by each actor with regard to the processing carried out by another actor. For example, who would be responsible for informing the data subjects and at what stage of the processing? This is also related to the *exercise of users' rights*, since the data subject can very hardly understand to whom a request should be issued and, moreover, problems may arise in case of controversies since the stakeholders may operate under different jurisdictions.

In a scenario where many controllers are involved and personal data are collected in different ways, trust (or *lack of trust*) is an important factor. The richer the user profile, the higher the temptation for the operators to target a user with unsolicited advertising or to engineer a pricing structure capable to extract as much surplus from the user as possible. This practice is known as price differentiation and the border between differentiation and discrimination may sometimes be very thin. The possibility of *data inference and re-identification* cannot be excluded unless appropriate technical safeguards are in place.

Moreover, in the context of an automated decision making process, the *risks of false positives* may be relevant, as well as the potential for externalities and social stigmata, as side-effect of the amplification triggered by the analysis of users personal data.

## A.4 Risk mitigation strategies

Following the aforementioned description, we provide a number of risk mitigation strategies that could be applied in the context of each of the smart cities big data analytics scenarios.

**Transparency and awareness**

Anticipating the rationale of the new General Data Protection Regulation, the different stakeholders may agree between themselves, and in compliance with the law, on the allocation of data protection responsibilities and on designating which of them plays the role of single point of contact (PoC) for privacy related issues, as well as for any technical assistance the user may need. This will facilitate enormously the exercise of users' rights and increase the effectiveness of the information. The creation of a web resource where the user can retrieve his/her own data may also be a valuable tool in this sense. For example in the case of citizen's platform, the application itself could offer the users' access to their posts and the uploaded images or videos, allowing also to correct and/or delete them permanently.

**User control**

User control needs to be reinforced, enabling users to effectively select between a number of choices (after having been informed on their impact on the user experience). For example, with specific regard to the parking application the forwarding of the mobile number from the telco to third parties (in order to offer added value services) should be selected by the used himself/herself. Tools such as privacy preferences and personal data stores can be very interesting in this regard, providing also for smoothness and ease of use (no need to "exit" or disrupt the service).

**Data minimization**

Also, in order to strengthen the data minimization principles, as much as possible the information handed over between the parties should be engineered in order to be Boolean rather than fully analytical. This is part of the data collection process and should be explicitly addressed in each different application. For example, in the smart parking application, instead of disclosing the entire amount of money available on the account for the payment, the binary variable YES/NO could be used. As another example, in the smart metering application, if the customer has a simple contract in which he/she pays the same price for electricity throughout the day, the meter may just collect a daily single reading and not a very detailed usage pattern. Also, if a user feels that his/her personal sphere is violated by the implementation of the metering devices, possibility should be left for the data subject to object to the installation of the smart meter.

**Access control and encryption**

In order to avoid crime related risks, access to data should be based on the respect of the principle of separation of duties, where each actor is enabled to have access to the data relating to the portion of the service that he provides, on a strict need-to-know basis. The relevant encryption based access controls and search mechanisms presented in Chapter 3 can be interesting solutions to explore in this context. For instance, in the smart metering scenario, holding much of the data in the meter in an encrypted format, allowing the provider to query them when needed for the provision of the service, can be a strong privacy guarantee. Other technical and organizational safeguards to minimize any risk of data misuse include log management, aggregation of data whenever individual level data are not required, audit and policy enforcement.

**Retention, deletion and anonymization**

In each of the presented use cases, the data should be kept only for the period that are absolutely necessary for the processing. For example, in the context of the smart parking application, personal data should not be retained after the user finds his/her preferred parking slot (unless he/she has requested additional services and has explicitly consented to this). Also, upon a data subject's specific request, and if no other legitimate interests or legally binding constraints exist, personal data should be deleted (e.g. in the citizen platform, webmasters should promptly remove the relevant piece of information from their website). If data need to be stored beyond the data retention periods, appropriate anonymization methods need to be applied.

**Privacy Impact Assessment**

Finally, a practical tool in this context, and more generally in all big data applications where many stakeholders are involved, is the use of a Privacy Impact Assessment (PIA) prior to the deployment of the service. In the specific context of smart grids, the European Commission has developed a PIA Template which can be usefully applied by network and smart grid system operators.

# Annex 2 – Anonymization background information

## A.5 On utility measures

In this section we give an overview of utility measures according to the state-of-the art, supporting the relevant topic in paragraph 4.1.1. They are measures to evaluate in what extent the application of anonymization procedures affects the outcome of inferences.

**Generic vs data-use specific utility measures**

If the data controller can anticipate the analyses that the users wish to carry out on the anonymized data, then he can choose SDC methods and parameters that, while adequately controlling disclosure risk, minimize the impact on those analyses. Unfortunately, the precise user analyses can seldom be anticipated by the data controller, let alone by the users themselves, when anonymized data are released for general use. On the other hand, releasing different anonymized versions of the same data set optimized for different data uses is a dangerous option, as it might result in data disclosure. Hence, SDC must often be based on generic measures.

**Generic utility measures for numerical data**

These measures are mostly intended for numerical attributes and they focus on the impact of anonymization on the statistics of the data set. Statistics to be examined include the values themselves, variances, correlations, covariances, and all statistics related to principal components (including communalities, factor score coefficients, correlations between attributes and principal components, etc.). The impact on a statistics can at least be measured in three ways: mean square error, mean absolute error or mean variation. See [192] for more details.

**Generic utility measures for categorical data**

If data are not numerical, we can resort to other procedures to measure the impact of anonymization on them:

- Direct comparison of the categorical values, to obtain some average distance between them (a distance for categorical values must be agreed upon).

- Comparison of contingency tables. Compute multidimensional contingency tables on both the original and the anonymized data sets and compute a distance between the two contingency tables.

- Entropy-based/probabilistic measures. These measure the uncertainty on the values of the original data set given the values of the anonymized data set.

See [192] for more details.

**Generic utility measures for any type of data**

In [193], several measures of information loss (i.e. utility loss) were proposed that compare the distribution of the original and the anonymized microdata. Among the proposed measures, the one based on propensity scores seems the most promising:

- Merge the original and anonymized data sets and add a binary attribute *T* taking value 1 for each anonymized record and value 0 for each original record.

- Regress *T* on the rest of attributes of the merged data set and call the adjusted attribute *T'*. Let the propensity score $p_i$ of record *i* of the merged data set be the value of *T'* for record *i*.

- Then utility is high if the propensity scores of the anonymized and the original records are similar; more specifically, if the following expression is close to 0: $U = \sum_{i=1}^{N}[p_i - \frac{1}{2}]^2$.

Certainly, the above measure captures whatever and probably no more utility than is captured by the regression model adjusted to the merged data set. For a fixed anonymization, the more sophisticated the regression model (i.e. the more higher-order interactions it includes), the more demanding is the utility metric *U* and the less likely it is to approach zero.

**Specific utility measures**

Specific utility measures correspond to measures tailored for particular uses of the data. In data mining, clustering and classification are the most usual types of data analysis, and can be applied to data with quite a different structure (from simple tables, to non-SQL databases and social networks). Information loss measures for clustering and classification have been defined. They compare the results of clustering and classification using the original data and the results of the same analysis using the protected file. In the case of clustering, measures compare the clusters obtained from original and protected files using partition distances based on e.g. Jaccard and Rand indices [194]. In the case of classification, measures compare accuracy and AUC (area under the curve) [195], [196] .

Naturally, other uses (and the corresponding information loss measures) more tailored to specific types of data have also been considered in the literature. See for example [197] about the case of association rule mining for market basket analysis.

These measures can be applied and have been used in the case of big data.

Note that in general, information loss measures are defined as follows

IL(X,X') = divergence(analysis(X), analysis(X'))

where the divergence is a distance on the space of the analysis.

## A.6 On disclosure risk measures, record linkage and attack models

There are Boolean and quantitative approaches for establishing an appropriate level of disclosure risk. Naturally, under a Boolean approach, files are either compliant or not with the privacy requirements. Under a quantitative approach, a numerical value expressing the degree of disclosure risk is calculated.
For Boolean approaches, data protection methods become a single objective optimization problem, as the goal is to maximize data utility (minimize information loss) given privacy constraints. For quantitative approaches, data protection methods become a multi objective optimization problem as the minimization of risk is, in general, in contradiction with the maximization of data utility.

Differential privacy (see 4.1.3.2) and k-anonymity (see 4.1.3.1) are examples of Boolean approaches. They establish a threshold over which a data release is considered safe. This level corresponds to the parameter ε in differential privacy and k in k-anonymity, or, in other words, we state that there is no disclosure risk in a data release when we have e.g. k-anonymity for a given k. Methods for differential privacy and k-anonymity are single objective optimization problems which focus on the minimization of information loss (given a convenient measure of information loss).

Note that k-anonymity is a definition focusing on identity disclosure, while differential privacy is on attribute disclosure.

For identity disclosure, measures of disclosure risk are based on the number of reidentifications for a released data file. This is usually estimated experimentally by means of record linkage, although other measures as uniqueness have also been defined.

- **Uniqueness:** It is defined [198] as the probability that rare combinations of attributes in the released and protected data set are also present in the population. Note that if such combinations exist in the population, we may have reidentification and disclosure.

- **Number of reidentifications**: The number or the proportion of records in a file that an intruder may be able to reidentify is a measure of its risk. Algorithms for data integration, in particular for schema and data matching, are used to evaluate the number of reidentifications. In particular, record linkage algorithms have been used for this purpose with the goal of linking the information available to the intruder with the released file. This approach has been used, for example, in [199], [200], [201] and [202].

k-Anonymity focuses on identity disclosure. It can be understood as a way to avoid re-identification and forcing that the probability of re-identification is at most 1/k. This link between k-anonymity and reidentification led to new definitions where the indistinguishability of k-records is replaced by conditions on linkability and reidentification. See e.g. k-concealment [203] [204], n-confusion [205], (k,t)-confusion [87].

Identity disclosure is evaluated in terms of the chances of finding an individual in the published file. Some privacy models as k-anonymity are implicitly based on this type of attacks. Reidentification and record linkage are the usual approaches to quantitatively evaluate this risk.

**Reidentification and record linkage models**

Record linkage models for disclosure risk assessment formalize and evaluate the scenario in which intruders attack a protected data set using the most effective programs with the goal of linking their information to this data set.

An advantage of reidentification approaches is that they can be used in a diversity of scenarios. We can use them to give numerical estimations of risk for synthetic data (see e.g. [201] & [202]) and also when the data published do not have the same variables (metadata/schema) that the data of the intruder (see e.g., [206] & [207]).

A difficulty is to know what type of information an intruder can access. Experiments for measuring disclosure risk using record linkage usually take subsets of the original file (or the complete original file) as the information of the intruder. However, other assumptions using less information have also been considered (e.g., [208] considers that the intruder knows r or less attributes). In the case of disclosure risk for graphs (and social networks) different types of attacks have been discussed in the literature (e.g., considering the degree of a node, considering a subgraph or considering that the intruder can only know the answers of a given query).

A formal model of reidentification algorithms is given in [209]. The model is based on the concept of true probability, which is the probability of reidentification when all information is available. Then, informally, reidentification algorithms should lead to a probability distribution that is compatible with the true probability, and the more information is available the closer is the distribution to the true probability. The model is based on imprecise probabilities and belief functions.

Formal models of reidentification algorithms are needed in order to properly define the worst-case scenario and to define upper bounds of disclosure risk. A formal approach of reidentification (via generalization) was also used to define k-concealment in [203] and [204].

**The worst-case scenario for re-identification**

The worst-case scenario for disclosure risk corresponds to the case in which the intruder has the most effective reidentification algorithm and the maximum information available related to the protected data file.

The information that may be available corresponds, at least, to the following

- **Information about the masking process**: That is, how data has been protected (masking method and its parameters).

- **Information on the respondents**: The most comprehensive information publicly available on the respondents in the file. Here, we understand with publicly available, information that can be somehow obtained (not necessarily free).

Several reidentification algorithms exist in the literature. Some of them are parametric. For example, distance-based record linkage requires a distance function to compare pairs of records (e.g. Euclidean, Mahalanobis, Bilinear forms, Choquet-integral based) and, in addition, the distance itself can depend on parameters (e.g. weights in weighted Euclidean and positive-definite matrices in bilinear forms).

The evaluation of this worst-case scenario needs to consider the best reidentification algorithm with the best parameterization. With respect to the information available to intruders, we can consider that they have the whole original file. We can use machine learning and optimization approaches to tune record linkage methods (find appropriate distances and parameters). For example, in [210] and [84] parameters for some distances were learnt. In this way, we can obtain upper bounds of the number of reidentification records.

Additional considerations can be made to further increase the effectiveness of attacks. For example, that the set of linked pairs is a one-to-one relationship between the original and the protected file [211].

## A.7 On anonymization methods

We have discussed in the document (paragraph 4.1.5) that there are three major ways of anonymizing a data set: perturbative, non-perturbative, and synthetic methods. We review these approaches below.

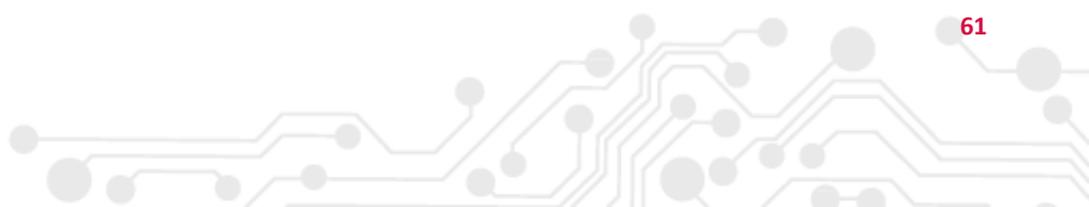

**Perturbative masking**

There are several principles for perturbative masking:

- **Noise addition**: This principle is only applicable to numerical microdata. The most popular method consists of adding to each record in the data set a noise vector drawn from a N(0,αΣ), with Σ being the covariance matrix of the original data. Means and correlations of original data can be preserved in the masked data by choosing the appropriate α. Additional linear transformations of the masked data can be made to ensure that the sample covariance matrix of the masked attributes is an unbiased estimator for Σ.

- **Microaggregation**: Microaggregation [212] partitions records in a data set into groups containing each at least k records; then the average record of each group is published. Groups are formed by the criterion of maximum within-group similarity: the more similar the records in a group, the less information loss is incurred when replacing them by the average record. There exist microaggregation methods for numerical and also categorical microdata.

- **Data swapping:** Values of attributes are exchanged among individual records, so that low-order frequency counts or marginals are maintained. Although swapping was proposed for categorical attributes, its rank swapping variant is also applicable to numerical attributes. In the latter, values of each attribute are ranked in ascending order and each value is swapped with another ranked value randomly chosen within a restricted range (e.g. the ranks of two swapped values cannot differ by more than p% of the total number of records).

- **Post-randomisation**: The PRAM method [213] works on categorical attributes: each value of a confidential attribute is stochastically changed to a different value according to a prescribed Markov matrix.

A form of noise addition was the anonymization method originally proposed in [47] to attain differential privacy. On the other hand, in [214], it was shown how microagreggating the quasi-identifier attributes of a data set yields k-anonymity.

**Non-perturbative masking**

Principles used for non-perturbative masking include:

- **Sampling**: Instead of publishing the original data file, only a sample of it is published. A low sampling fraction may suffice to anonymize categorical data (probability that a sample unique is also a population unique is low). For continuous data, sampling alone does not suffice.
- **Generalization**: This principle is also known as coarsening or global recoding. For a categorical attribute, several categories are combined to form new (less specific) categories; for a numerical attribute, numerical values are replaced by intervals (discretization).

- **Top/bottom coding**: Values above, resp. below, a certain threshold are lumped into a single top, resp. bottom, category.

- **Local suppression:** Certain values of individual attributes are suppressed in order to increase the set of records agreeing on a combination of quasi-identifier attributes. This principle can be combined with generalisation.

The original approach proposed in [42] to attain k-anonymity consisted in generalization combined with local suppression.

**Synthetic microdata generation**

Rubin [215] proposed *fully synthetic data generation*, which consists of randomly generating data in such a way that some statistics or relationships of the original data are preserved. The advantage of fully synthetic data is that no respondent re-identification seems possible, because data are artificial. There are downsides, too. If a synthetic record matches by chance a respondent's attributes, re-identification is likely and the respondent will find little comfort in the data being synthetic. Data utility of synthetic microdata is limited to the statistics and relationships pre-selected at the outset: analyses on random subdomains are no longer preserved. Partially synthetic or hybrid data are more flexible.

There are two approaches seeking to reap the best of synthetic and original data:

- **Partially synthetic data** [216]: Only the most sensitive data (i.e. the values of some variables and/or records) are replaced by synthetic versions before release. For the rest, original data are released.

- **Hybrid data** (e.g. [217], [218]: Original and synthetic data are combined and the combination yields so-called hybrid data, which are the data subsequently released in place of original data. Depending on how the combination is computed, hybrid data can be closer to the original data or to the synthetic data.

Partial and hybrid data are the only forms of synthetic data that can be used for anonymization at source if one wants to create big data by linking anonymized coming from several sources. Extensive background on synthetic data generation can be found in [219].

**The permutation paradigm of anonymization and comparability**

Let $X = \{x_1, x_2, \ldots, x_n\}$ be the values taken by attribute *X* in the original data set. Let $Y = \{y_1, y_2, \ldots, y_n\}$ be the anonymized version of *X*. We make no assumptions about the anonymization method used to generate *Y*, but we assume that the values of *X* and *Y* can be ranked in some way; any ties in them are broken randomly. Knowledge of *X* and *Y* allows deriving another set of values $Z = \{z_1, z_2, \ldots, z_n\}$ using the following reverse-mapping procedure given in [220]:

**for** i=1 **to** n: compute $j = Rank(y_i)$ and set $z_i = x_{(j)}$; **endfor**

where $x_{(j)}$ is the value of *X* of rank *j*. Note that the reverse-mapped attribute *Z* is a permutation of *X*.

Since the previous reverse-mapping procedure is applicable to any anonymization method, we can say that *any anonymization method is functionally equivalent to a two-step procedure consisting of a permutation step* (mapping the original data set **X** to the data set **Z** obtained by running the reverse-mapping procedure for all attributes) *plus a residual noise addition step* (adding the difference between **Z** and the anonymized **Y**). We say the noise addition is necessarily residual (small) because it cannot entail any change in ranks (**Z** and **Y** have identical ranks, by construction of the reverse-mapping procedure).

In this light, it seems rather obvious that protection against re-identification via record linkage comes from the permutation step in the above functional equivalence. Thus, *any two anonymization methods can, however different their actual operating principles, be compared in terms of how much permutation they achieve, that is, how much they modify ranks* [220]. Thus, the permutation paradigm makes comparability easy.

**The permutation paradigm and anonymization verifiability by subjects**

Beyond making comparability of methods easy, the above permutation paradigm makes it possible for the subject to verify the level of anonymization of her record. Since what basically counts in anonymization is permutation, he/she just needs to check the amount of permutation that the values in him/her original record have undergone in the anonymized record.

To check permutation, the data subject looks for the maximum distance *d* such that there is no anonymized record such that, *for every attribute*, the rank of the attribute value differs less than *d* from the rank that the attribute value in the original record would have if inserted in the anonymized data set. Call *d* the *permutation distance* for the subject's record. Additionally to checking the permutation level, the subject can check that anonymized values within distance *d* of her original attribute values are diverse enough.

**The permutation paradigm and a maximum-knowledge adversary**

Estimating the adversary's background knowledge is a thorny issue, especially in a big data context. Hence, we need a definition of adversary that is independent of the background knowledge. To that end, we can draw inspiration from cryptography. The following attacks are the usual ones in cryptography: ciphertext-only (the adversary knows only the ciphertext), known-cleartext (the adversary knows some pairs cleartext-ciphertext), chosen-cleartext (the adversary can get the ciphertext corresponding to any cleartext of his choice) and chosen-ciphertext (the adversary can get the cleartext corresponding to any ciphertext of his choice).

If we move from cryptography to anonymization, we can equate the original data set to cleartext and the anonymized data set to ciphertext. If we focus on anonymization of data sets, interactive attacks (such as chosen-cleartext and chosen-ciphertext) do not make sense any more.

Hence, the strongest attack we need to consider is known-cleartext. In this attack, one might think of an intruder knowing particular original records and their corresponding anonymized versions; however, this is unlikely, because anonymization precisely breaks the links between anonymized records and corresponding original records. A more reasonable definition for a *known-cleartext attack in anonymization is one in which the adversary knows the entire original data set (cleartext) and the entire corresponding anonymized data set (ciphertext), his objective being to recreate the mapping between the original and the anonymized records.*

The above definition of adversary is stronger than any other prior such definition in the data set anonymization scenario. In particular, assuming the adversary knows the entire original data set *makes any adversary's background knowledge irrelevant*. Note also that our adversary is purely malicious: he does not gain any new knowledge by his attack (because he already knows the entire original data set); however, he may seek to tarnish the reputation of the data controller by boasting the mapping between original and anonymized records. In fact, such a maximum-knowledge adversary knowing **X** and **Y** can use the above reverse-mapping procedure to get **Z**. He then can use **Z** instead of **Y**, so that his problem becomes finding the permutation between **X** and **Z**.

Finally, note that our maximum-knowledge adversary makes any deterministic anonymization method insecure: the adversary can apply the deterministic method to **X** and determine the mapping between **X** and **Z**. Hence, only randomized anonymization methods can withstand our powerful adversary.

**The permutation paradigm and transparency to users**

Transparency to the data user means giving the user all anonymization details, except any random seeds used for pseudo-randomization by the anonymization method. Withholding the anonymization parameters is problematic: (i) the legitimate user cannot properly evaluate the utility of the anonymized data; (ii) basing disclosure protection on the secrecy of the anonymization parameters is a poor idea (it violates Kerckhoff's principle). Being transparent is good for the data user and is of no consequence to the other stakeholders in the anonymization process if we have a maximum-knowledge attacker:

- The subject can verify how permuted have been the values in his/her record (see above); this is all that matters to him/her, regardless of whether the data user has been given information on the anonymization algorithm.

- Our maximum-knowledge adversary can perform record linkage between X and Z without using any information on the anonymization method; to him/her, anonymization is just a permutation that he/she seeks to recreate. A rational procedure is for the adversary to link each record in X to the record in Z at minimum permutation distance.

See [220] for more details on the permutation paradigm and its implications for comparability, verifiability, background knowledge and transparency and on how the adversary can check the plausibility of his record linkages.

## A.8 Transparency and data privacy

Publishing data together with information on how this data has been processed is a good practice. Transparency in data privacy pursues this goal. Publishing how data has been anonymized increases data utility but also disclosure risk. At the same, we can consider the development of methods that are resistant to transparency attacks. We detail these issues below.

**Transparency attacks**

The effects of transparency can be evaluated using reidentification algorithms. To this end, specific algorithms have been developed. They take into account information about the masking method and also, if any, their parameters.

Specific algorithms for data protected using rank swapping [221] and microaggregation [222], [221] and [223] follow the same approach. In short, rank swapping and microaggregation are applied, respectively, attribute-wise or to subsets of attributes independently. Then, given data from an individual the intruder can build an anonymity set (a set of possible links) for each attribute or set of attributes. Given an intruder's record, it is sure that this record is in the intersection of these anonymity sets. In addition, when the set contains a single record, we have reidentification and, thus, identity disclosure. This type of procedure can be seen as an intersection attack.

An important difference between specific record linkage and generic record linkage is that in the former the intruder may arrive to cases where the intersection set is a single record, and, thus, be sure that the attack has been effective. In generic record linkage, we get an estimation of the number of reidentifications obtained by the intruder but the intruder is not always able to distinguish those correct links (real matches) and those that are not correct.

For rank swapping, the transparency attack is quite effective, as with a few variables the attack already results into several intersection sets with a single record. For optimal univariate microaggregation similar results apply. For heuristic multivariate microaggregation, this type of attack is not so effective.

Nevertheless, heuristic methods without randomness, as MDAV, are reproducible and when the intruder has the whole original file, he can reidentify the records.

**Transparency attack-safe masking**

When data are published under the transparency principle, intruders can use specific record linkage algorithms to attack the data. Because of that, methods have to be resistant to this type of attacks.

There are studies on the transparency risk of noise addition, rank swapping, and microaggregation. Also, the amount of information released can imply different risks. For example [224] states that there is not much difference on the disclosure risk of only publishing data and publishing data with the covariance matrix of the error added. However, the release of additional parameters increases the risk.

Some rank swapping variants have been developed to be transparency-safe. This is the case of rank swapping p-buckets and rank swapping p-distribution [221]. They avoid the transparency attack ensuring that the intersection set is the whole database. Fuzzy microaggregation was introduced with a similar goal. See [225] and [226] for details.

**Transparency attacks**

The effects of transparency can be evaluated using reidentification algorithms. To this end, specific algorithms have been developed. They take into account information about the masking method and also, if any, their parameters.

Specific algorithms for data protected using rank swapping [221] and microaggregation [222], [221] and [223] follow the same approach. In short, rank swapping and microaggregation are applied, respectively, attribute-wise or to subsets of attributes independently. Then, given data from an individual the intruder can build an anonymity set (a set of possible links) for each attribute or set of attributes. Given an intruder's record, it is sure that this record is in the intersection of these anonymity sets. In addition, when the set contains a single record, we have reidentification and, thus, identity disclosure. This type of procedure can be seen as an intersection attack.

An important difference between specific record linkage and generic record linkage is that in the former the intruder may arrive to cases where the intersection set is a single record, and, thus, be sure that the attack has been effective. In generic record linkage, we get an estimation of the number of reidentifications obtained by the intruder but the intruder is not always able to distinguish those correct links (real matches) and those that are not correct.

For rank swapping, the transparency attack is quite effective, as with a few variables the attack already results into several intersection sets with a single record. For optimal univariate microaggregation similar results apply. For heuristic multivariate microaggregation, this type of attack is not so effective. Nevertheless, heuristic methods without randomness, as MDAV, are reproducible and when the intruder has the whole original file, he can reidentify the records.

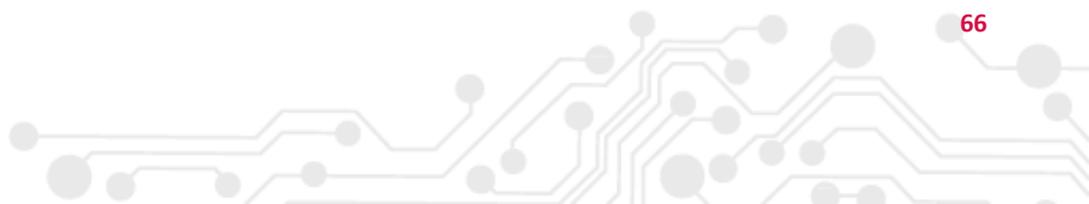

difference on the disclosure risk of only publishing data and publishing data with the covariance matrix of the error added. However, the release of additional parameters increases the risk.

Some rank swapping variants have been developed to be transparency-safe. This is the case of rank swapping p-buckets and rank swapping p-distribution [221]. They avoid the transparency attack ensuring that the intersection set is the whole database. Fuzzy microaggregation was introduced with a similar goal. See [225] and [226] for details.

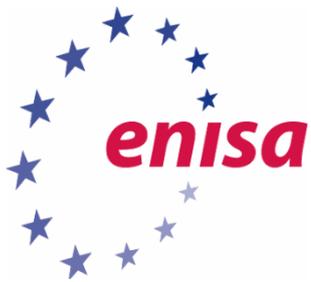